\documentclass[twocolumn,showpacs,preprintnumbers,amsmath,amssymb,superscriptaddress]{revtex4}
\usepackage{graphicx,dcolumn,bm}

\newcommand{\be}{\begin{equation}}
\newcommand{\ee}{\end{equation}}
\newcommand{\bea}{\begin{eqnarray}}
\newcommand{\eea}{\end{eqnarray}}
\newcommand{\bsube}{\begin{subequations}}
\newcommand{\esube}{\end{subequations}}
\newcommand{\Sec}[1]{Sec.\,\ref{#1}}
\newcommand{\Eq}[1]{Eq.\,(\ref{#1})}
\newcommand{\Eqs}[1]{Eqs.\,(\ref{#1})}
\newcommand{\Fig}[1]{Fig.\,\ref{#1}}

\newcommand{\rmS}{{\rm S}}
\newcommand{\rmB}{{\rm B}}
\newcommand{\rmT}{{\rm T}}

\newcommand{\rmi}{{\rm i}}
\newcommand{\rmd}{{\rm d}}

\newcommand{\bmchi}{{\bm{\chi}}}

\newcommand{\ra}{\rangle}
\newcommand{\la}{\langle}

\newcommand{\sgm}{\sigma}
\newcommand{\Omg}{\Omega}
\newcommand{\omg}{\omega}
\newcommand{\Gam}{\Gamma}
\newcommand{\gam}{\gamma}
\newcommand{\Dlt}{\Delta}
\newcommand{\dlt}{\delta}
\newcommand{\lmd}{\lambda}
\newcommand{\vpl}{\varepsilon}
\newcommand{\epl}{\epsilon}
\newcommand{\upa}{\uparrow}
\newcommand{\dwa}{\downarrow}
\newcommand{\Lmd}{\Lambda}

\newcommand{\tr}{{\rm tr}}
\newcommand{\gamRF}{\gamma_{_{\rm RF}}}

\newcommand{\hfOmg}{{\textstyle \frac{\Omega}{2}}}

\begin{document}

\title{Stochastic thermodynamics of an electron-spin-resonance quantum dot system}

\author{JunYan Luo}\email{jyluo@zust.edu.cn}
\affiliation{Department of Physics, Zhejiang University of Science
  and Technology, Hangzhou 310023, China}
\author{Yiying Yan}
\affiliation{Department of Physics, Zhejiang University of Science
  and Technology, Hangzhou 310023, China}
\author{Hailong Wang}
\affiliation{Department of Physics, Zhejiang University of Science
  and Technology, Hangzhou 310023, China}
\author{Jing Hu}
\affiliation{Department of Physics, Zhejiang University of Science
  and Technology, Hangzhou 310023, China}
\author{Xiao-Ling He}
\affiliation{Department of Physics, Zhejiang University of Science
  and Technology, Hangzhou 310023, China}
\author{Gernot Schaller}
\affiliation{Institut f\"{u}r Theoretische Physik, Technische
Universit\"{a}t Berlin, Hardenbergstrasse 36, D-10623 Berlin, Germany}

\date{\today}

\begin{abstract}
We present a stochastic thermodynamics analysis of an electron-spin-resonance
pumped quantum dot device in the Coulomb--blocked regime, where a pure spin
current is generated without an accompanying net charge current.
Based on a generalized quantum master equation beyond secular
approximation, quantum coherences are accounted for in terms of an
effective average spin in the Floquet basis.
Elegantly, this effective spin undergoes a precession about an effective magnetic
field, which originates from the non-secular treatment and energy renormalization.
It is shown that the interaction between effective spin and effective
magnetic field may have the dominant roles to play in both energy transport and
irreversible entropy production.
In the stationary limit, the energy and entropy balance relations are also
established based on the theory of counting statistics.
\end{abstract}

\maketitle

\section{\label{thsec1}Introduction}

The state-of-the-art nanofabrication is able to create small
systems far from the thermodynamic
limit, where both thermal and quantum mechanical fluctuations
have essential roles to play.
This opens up opportunities to create new functional devices, but
also poses great challenges to manipulate nanoscale systems,
which interact with their environments and exchange energy in a random
manner \cite{Wei08}.
Understanding thermodynamics from quantum
mechanics \cite{Gem10,Sek10,Cam11771,Sei12126001,Esp15080602,Vin16545,And17010201,%
Car16240403,Ali180108314,Ben171} is thus
of fundamental significance to characterize energy fluctuations at the
microscopic level, and also of technological importance for the design of
efficient quantum heat engines \cite{Kos14365,Uzd15031044,Leg16022122,Ros16325,%
Arg17052106,Rou17062131,Rei1760006,Sco18062121,Hew18042102} and exploration of
information processing capabilities \cite{Goo16143001,Esp091665,Bro10937,%
Ber12187,Esp1230003,Bar14090601,Par15131,Gol17010601,Str17021003,Ito18030605}.

%%%%%%%%%%%%%%%%%%%%%%%%%%%%%%%%%%%%%%%%%%%%%%%%%%%%%%%%%%%%%%%%%%%%%%%%%
In comparison with a soft-matter system, where fluctuation relations were
first measured experimentally \cite{Toy10988}, a solid-state device is
considered to be an ideal testbed to investigate thermodynamics of open
quantum systems due to a number of intriguing advantages \cite{Pek15118,Pek19193}.
For instance, solid-state systems are robust such that experiments can be
repeated normally up to million times under the same condition.
Moreover, the particles and quantum states, as well as their couplings
to the environments can be manipulated in a precise way.
Measurement of the statistics of the dissipated energy is proved
to satisfy the Jarzynski equality and Crooks fluctuation
relations \cite{Sai12180601}.
Furthermore, the generalized Jarzynski equality has also been
validated in a double-dot Szilard engine under feedback
control \cite{Kos14030601}.
So far, experimentalists have been able to implement a quantum Maxwell demon
either in a superconducting circuit \cite{Cot1704827,Mas181291} or in a
single electron box \cite{Koski1413786},
where the intimate relation between work and
information is unambiguously revealed.

Recent progress in solid-state engineering has made it possible to control
spin coherence to the timescale of seconds \cite{Bar131743,Par17372},
ushering thus in a new era of ultracoherent
spintronics \cite{Zut04323,Aws131174}.
This provides an exciting opportunity to incorporate spintronics into
thermodynamics and evaluate both energetic and entropic costs to
manipulate spin information.
In contrast to a conventional electronic setup, where information
and energy are transmitted via charge,
in a spintronic device it
is the spin that will work as a vehicle for energy and information
transduction.
However, as an intrinsic angular momentum, spin is not conserved
in general.
It is therefore essential to explore the energy and entropy balance
relations in terms of a pure spin current without accompanying
a charge current and understand what kind of roles the dynamics
of spin will play in these processes.

This work is devoted to unveil the underlying mechanisms by
analyzing the stochastic thermodynamics of an electron-spin-resonance
(ESR) pumped quantum dot (QD) system in the Coulomb--blockaded regime,
where a pure spin current is generated without a net charge flow.
Based on a generalized quantum master equation (GQME) beyond secular
approximation, the effect of coherences
is taken into account via an effective average spin,
which builds up and decays due to tunnel coupling to an electrode.
Remarkably, the non-secular treatment and energy renormalization
give rise to an effective magnetic field, about which the
effective spin undergoes precession.
It is revealed unambiguously that the interaction between effective spin
and effective magnetic field may play the dominant roles
in energy flow as well as in the irreversible entropy production.

This paper is organized as follows. We begin in
\Sec{thsec2} with an introduction of the ESR pumped QD system, where
a pure spin current is generated without accompanying a net charge
current.
The GQME is derived in \Sec{thsec3}, with special attention paid to
the unique influence of non-secular treatment on the spin dynamics
and spin current.
In particular, the quantum coherences are revealed to have vital
roles to play in energy current.
\Sec{thsec4} is devoted to the analysis of stochastic thermodynamics
based on the counting statistics, where both energy balance and
entropy balance relations are established.
Furthermore, the influence of the interaction between effective
spin and magnetic field on irreversible entropy production is revealed.
Finally, we summarize the work in \Sec{thsec5}.

\section{\label{thsec2}Model Description}

\begin{figure}
\includegraphics*[scale=0.95]{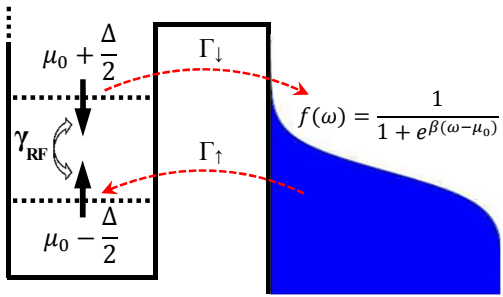}
\caption{\label{Fig1}
A schematic of an ESR pumped QD system, where a single QD is tunnel coupled
to a side electron reservoir characterized by the Fermi function $f(\omg)$,
with inverse temperature $\beta=(k_\rmB T)^{-1}$ and chemical potential $\mu_0$
located in the middle of the spin-up ($\mu_0-\frac{\Dlt}{2}$) and spin-down
($\mu_0+\frac{\Dlt}{2}$) levels.
The ESR pumping produces a pure spin current in the absence of a net charge
current.}
\end{figure}

We investigate an ESR pumped system closely related to
experiments \cite{Xia04435,Elz04431}.
The system is comprised of a Coulomb--blockaded single QD,
tunnel coupled to a side electron reservoir (\Fig{Fig1}).
The QD is exposed to a local external rotating magnetic field
\be\label{Borg}
\bm{B}=\{B_\parallel \cos\Omg t, B_\parallel \sin\Omg t, B_\perp\},
\ee
where its $z$ component leads to the Zeeman splitting of
the single level $\Dlt=g_{\rm z}\mu_{\rm B}B_\perp$, with
$g_{\rm z}$ the electron $g$-factor in the z direction and
$\mu_{\rm B}$ the Bohr magneton.
The $x$ and $y$ components of the magnetic field are oscillating
in time, where the frequency $\Omg$ is tuned very
close to $\Dlt$, resulting in the well-known ESR and spin flipping in QD.
The electron spin is further tunnel coupled to a side
reservoir, whose chemical potential $\mu_0$ is set in the middle
of the split spin-up and spin-down levels.
A spin-up electron may tunnel into QD, where it is pumped to the higher
level with its spin orientation flipped and finally tunnels out
to the side reservoir. At sufficiently low temperatures, this generates
an ESR-pumped spin current without accompanying a net charge current.
The Hamiltonian of the total (T) system reads
\be\label{Htot}
H_\rmT(t)=H_\rmS(t)+H_\rmB+V.
\ee
The first term describes the QD with ESR pumping
\begin{align}\label{HQD}
H_\rmS(t)=&\frac{\Dlt}{2}(d^\dag_\dwa d_\dwa-d^\dag_\upa d_\upa)
+U_{\rm C}  d^\dag_\upa d_\upa d^\dag_\dwa d_\dwa
\nonumber \\
&+\gamRF(d^\dag_\upa d_\dwa e^{\rmi \Omg t}+d^\dag_\dwa d_\upa
e^{-\rmi \Omg t}),
\end{align}
where $d^\dag_{\sgm}$ and $d_\sgm$ are the creation and annihilation
operators of an electron with spin $\sgm=\{\upa,\dwa\}$ in the QD.
Spin-up and spin-down states are coupled to each other due to the
rotating magnetic field, with the ESR Rabi frequency given by
$\gamRF=g_\parallel \mu_{\rm B}B_\parallel$ and $g_\parallel$ the electron
$g$-factor perpendicular to z.
We stress that in the following we will consider the Coulomb-blockaded limit
$U_{\rm C}\rightarrow\infty$,
thus effectively forbidding the double occupancy of the QD.

The second term in \Eq{Htot} depicts the side electron reservoir,
which is modeled as a collection of noninteracting electrons
\be\label{HB}
H_\rmB=\sum_\sgm H_\rmB^{(\sgm)}=\sum_{\sgm}
\left\{\sum_k \vpl_{k\sgm}c^\dag_{k\sgm}c_{k\sgm}\right\},
\ee
where $H_\rmB^{(\sgm)}$ is defined implicitly, with $c^\dag_{k\sgm}$ ($c_{k\sgm}$)
the creation (annihilation) operator for an electron with momentum $k$ and
spin $\sigma$.
For later use, we also introduce the operator for the number of spin-$\sgm$
electrons in the electron reservoir
\be\label{NB}
N_\rmB^{(\sgm)}=\sum_k c^\dag_{k\sgm}c_{k\sgm},
\ee
such that the operator for the total number of electrons in the reservoir
is $N_\rmB=\sum_\sgm N_\rmB^{(\sgm)}$.
The electrode is assumed to be in equilibrium, so that it can be
characterized by the Fermi distribution $f(\omg)=\{1+e^{\beta (\omg-\mu_0)}\}^{-1}$,
with the inverse temperature $\beta=(k_\rmB T)^{-1}$ and the chemical
potential $\mu_0$ set in the middle of spin-up and spin-down
levels.

The last term in \Eq{Htot} stands for tunnel coupling between
the single QD and side reservoir
\begin{align}\label{Hprime}
V=\sum_{\sigma}(f_{\sigma}^\dag d_{\sigma}+d_\sigma^\dag f_{\sigma}),
\end{align}
where $f_{\sgm}\equiv\sum_k t_{k\sgm} c_{k\sgm}$, with $t_{k\sgm}$ the
spin-dependent tunneling amplitude.
The corresponding tunneling rate for an electron with spin $\sgm$ is
characterized by the intrinsic linewidth
$\Gam_{\sgm}(\omg)=2\pi\sum_k|t_{k\sgm}|^2\dlt(\omg-\vpl_{k\sgm})$.
Hereafter, we assume wide band limit in the electrodes, which leads to energy
independent tunneling rates $\Gam_{\sgm}(\omg)=\Gam_{\sgm}$.
% %%
In what follows, we set unit of $\hbar=e=1$ for the Planck constant
and electron charge, unless stated otherwise.

\section{\label{thsec3}GQME and Spin Dynamics}

In order to describe the energy and particle transport between the QD and
side reservoir, we employ the GQME by including corresponding counting fields.
Assume the entire system is initially factorized and can be described by the
density matrix $\rho_\rmT(0)=\rho(0)\otimes\rho_\rmB$, where $\rho(0)$ is
the density matrix of the reduced system at time $t=0$ and $\rho_\rmB$ is
that of side reservoir at equilibrium.
The generalized density matrix evolves according to \cite{Bla001,Naz03}
\begin{widetext}
\be\label{rhoTchi}
\rho_\rmT(\bmchi,t)=\exp_+\left\{-\rmi\int_0^t \rmd\tau H_\rmT(\bmchi,\tau)\right\}\rho_\rmT(0)
\exp_-\left\{\rmi\int_0^t \rmd\tau H_\rmT(-\bmchi,\tau)\right\},
\ee
where the subscripts $+$ and $-$ stand for time ordering and anti-ordering,
respectively. The $\bmchi$-dressed Hamiltonian for the entire system in
\Eq{rhoTchi} is given by
\begin{align}\label{Hpchi}
H_\rmT(\bmchi,t)=&\exp\left\{\frac{\rmi}{2} \sum_{\sgm}
(\chi_{1\sgm} N^{(\sgm)}_\rmB+\chi_{2\sgm} H_\rmB^{(\sgm)})\right\}
H_\rmT(t)
\exp\left\{-\frac{\rmi}{2} \sum_{\sgm}
(\chi_{1\sgm} N^{(\sgm)}_\rmB+\chi_{2\sgm} H_\rmB^{(\sgm)})\right\},
\end{align}
\end{widetext}
where $H_\rmB^{(\sgm)}$ and $ N^{(\sgm)}_\rmB$ are defined in \Eqs{HB} and
(\ref{NB}), respectively. The involved counting fields read
$\bmchi=\{\chi_{1\upa},\chi_{1\dwa},\chi_{2\upa},\chi_{2\dwa}\}$, with
$\chi_{1\sgm}$ the counting field related to
the tunneling of spin $\sgm$ electron, and $\chi_{2\sgm}$ for energy
transmission associated with $\sgm$ spin.
Since the operators for the QD and reservoir commute, \Eq{Hpchi} can be
readily expressed as
\be
H_\rmT(\bmchi,t)=H_\rmS(t)+H_\rmB+V(\bmchi),
\ee
where $H_\rmS(t)$ and $H_\rmB$ remain unchanged.
It is only the tunnel-coupling Hamiltonian that becomes counting fields
dependent
\begin{align}\label{ARB}
V(\bmchi)&=\sum_{\sigma}\{f_{\sigma}(\bmchi) d^\dag_{\sigma}+{\rm h.c.}\},
\end{align}
with
$f_{\sgm}(\bmchi)=\sum_{k}t_{k\sgm}c_{k\sgm}
e^{-\frac{\rmi}{2}(\chi_{1\sgm}+\chi_{2\sgm}\vpl_{k\sgm})}$.

If one were given $\rho_{\rm T}(\bmchi,t)$ in \Eq{rhoTchi},
the reduced density matrix in the $\bmchi$ space can be readily obtained
via $\rho(\bmchi,t)=\tr_{\rm B}\{\rho_{\rm T}(\bmchi,t)\}$, where
$\tr_{\rm B}\{\cdots\}$ represents the trace over the degrees
of freedom of reservoir.
It leads directly to the cumulant generating function (CGF) in the
steady state
\be
{\cal F}(\bmchi)=\lim_{t\rightarrow\infty}\frac{1}{t}
\ln\{\tr_\rmS[\rho(\bmchi,t)]\},
\ee
where $\tr_\rmS[\cdots]$ means trace over the degrees of freedom
of the reduced system.
The statistics of particle and energy currents can be evaluated by
simply taking derivatives of the CGF with corresponding counting
fields.
However, we are neither able to nor interested to track the
dynamical evolution of the full system-plus-environment. Instead,
we would like to describe the reduced system by a dynamical
equation that accounts for (usually approximately) the influence
of the environment on the system state, while removing the need
to track the full environment evolution.

\begin{figure*}
\includegraphics*[scale=0.93]{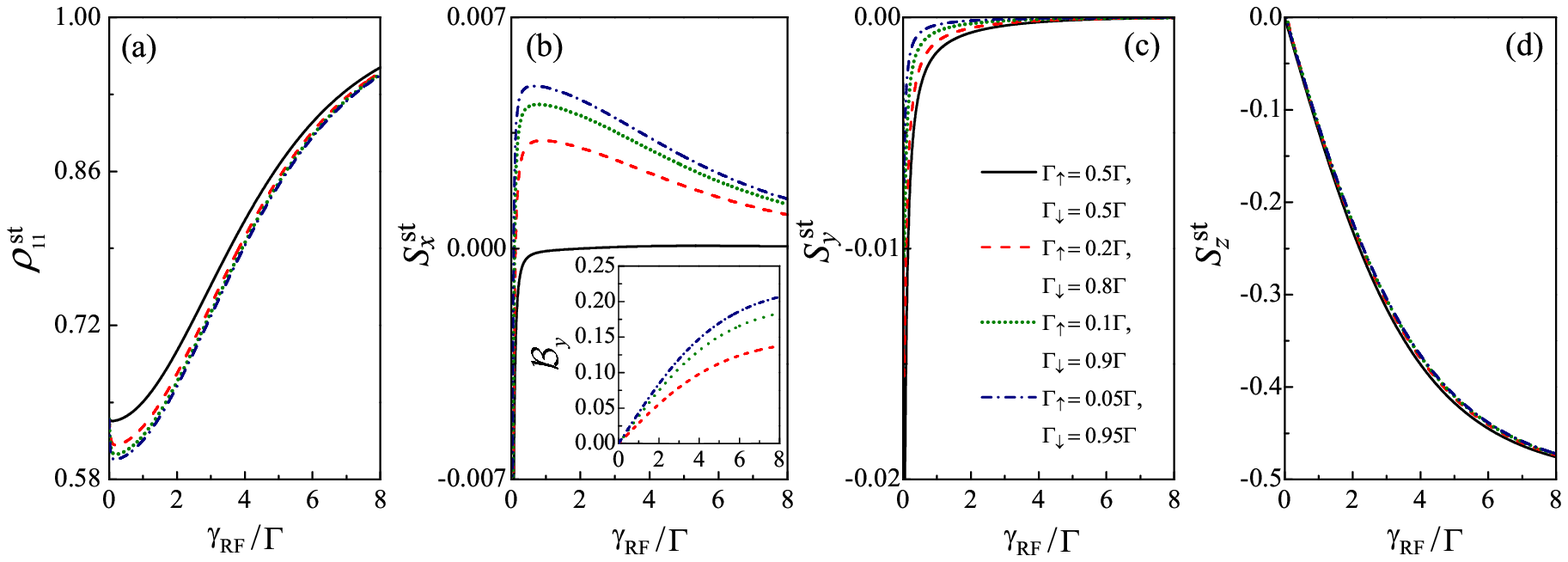}
\caption{\label{Fig2}
Stationary occupation of the QD ($\rho^{\rm st}_{11}$) and
accumulation of effective spin ($S_x^{\rm st}$, $S_y^{\rm st}$, $S_z^{\rm st}$)
versus Rabi frequency for different configurations of asymmetry in spin tunneling.
We set $\Gam=\Gam_\upa+\Gam_\dwa$ as the reference of energy.
Other plotting parameters are $\dlt=0$, $\Omg/\Gam=3.0$, $\beta \Gam=0.1$, and
a wide bandwidth ${\rm w}/\Gam=100$.
The inset in \Fig{Fig2}(c) shows the $y$-component of the effective magnetic
field vs. Rabi frequency.}
\end{figure*}

Here, we assume that the system and reservoir are weakly coupled and
perform a second--order perturbation expansion in terms of the
coupling Hamiltonian. It is then followed by the conventional
Born-Markov approximation but {\it without} invoking the widely
used secular approximation \cite{Hon09051129}.
To deal with the time-dependent system Hamiltonian, we work in the
Floquet basis \cite{Shi65B979,Hol941950,Gri98229,Szc13012120}.
The GQME finally reads (A detailed derivation
is referred to Appendix \ref{App-A})
\begin{align}\label{ZZD}
\dot{\rho}(\bmchi,t)=-\rmi[H_\rmS(t),\rho(\bmchi,t)]+{\cal R}(\bmchi)\rho(\bmchi,t),
\end{align}
where the first term describes the free evolution. The
second term stands for dissipation
\begin{widetext}
\begin{align}\label{ZZC}
{\cal R}(\bmchi)\rho(\bmchi,t)
	=&\bigg\{\Gamma_{0,+}(\bmchi){\cal J}[|u_0(t)\ra\la u_+(t)|]
	-\frac{1}{2}\Gamma_{0,+}{\cal A}[|u_0(t)\ra\la u_+(t)|]
	-\frac{\rmi}{2}\kappa_{0,+}{\cal C}[|u_0(t)\ra\la u_+(t)|]	
	\bigg\}{\rho}(\bmchi,t)
	\nonumber \\
	%%%
	&+\bigg\{\Gamma_{0,-}(\bmchi){\cal J}[|u_0(t)\ra\la u_-(t)|]
	-\frac{1}{2}\Gamma_{0,-}{\cal A}[|u_0(t)\ra\la u_-(t)|]
	-\frac{\rmi}{2}\kappa_{0,-}{\cal C}[|u_0(t)\ra\la u_-(t)|]	
	\bigg\}{\rho}(\bmchi,t)
	\nonumber \\
	%%%
	&+\bigg\{\Gamma_{+,0}(\bmchi){\cal J}[|u_+(t)\ra\la u_0(t)|]
	-\frac{1}{2}\Gamma_{+,0}{\cal A}[|u_+(t)\ra\la u_0(t)|]
	-\frac{\rmi}{2}\kappa_{+,0}{\cal C}[|u_+(t)\ra\la u_0(t)|]		
	\bigg\}{\rho}(\bmchi,t)
	\nonumber \\
	%%%
	&+\bigg\{\Gamma_{-,0}(\bmchi){\cal J}[|u_-(t)\ra\la u_0(t)|]
	-\frac{1}{2}\Gamma_{-,0}{\cal A}[|u_-(t)\ra\la u_0(t)|]
	-\frac{\rmi}{2}\kappa_{-,0}{\cal C}[|u_-(t)\ra\la u_0(t)|]
	\bigg\}{\rho}(\bmchi,t)
	\nonumber \\
	%%%
	&-\{[\Upsilon_{0,-}+\rmi \xi_{0,-}] |u_+(t)\ra\la u_-(t)| {\rho}(\bmchi,t)
	+[\Upsilon_{0,+}-\rmi \xi_{0,+}] {\rho}(\bmchi,t)
	|u_+(t)\ra\la u_-(t)|\}
	\nonumber \\
	&-\{[\Upsilon_{0,+}+\rmi \xi_{0,+}]
	|u_-(t)\ra\la u_+(t)|{\rho}(\bmchi,t)
	+[\Upsilon_{0,-}-\rmi \xi_{0,-}] {\rho}(\bmchi,t) |u_-(t)\ra\la u_+(t)|\}
	\nonumber \\
	&+	[\Upsilon_{+,0}(\bmchi)+\rmi \xi_{+,0}(\bmchi)
	+\Upsilon_{-,0}(\bmchi)-\rmi \xi_{-,0}(\bmchi)]
	|u_+(t)\ra\la u_0(t)|{\rho}(\bmchi,t)	|u_0(t)\ra\la u_-(t)|
	\nonumber \\
	&+[\Upsilon_{0,+}(\bmchi)-\rmi \xi_{0,+}(\bmchi)
	+\Upsilon_{0,-}(\bmchi)+\rmi \xi_{0,-}(\bmchi)] |u_0(t)\ra\la u_-(t)|{\rho}(\bmchi,t)	|u_+(t)\ra\la u_0(t)|
	\nonumber \\
	&+[\Upsilon_{+,0}(\bmchi)-\rmi \xi_{+,0}(\bmchi)
	+\Upsilon_{-,0}(\bmchi)+\rmi \xi_{-,0}(\bmchi)]
	|u_-(t)\ra\la u_0(t)|{\rho}(\bmchi,t)	|u_0(t)\ra\la u_+(t)|
	\nonumber \\	
	&+[\Upsilon_{0,+}(\bmchi)+\rmi \xi_{0,+}(\bmchi)
	+\Upsilon_{0,-}(\bmchi)-\rmi \xi_{0,-}(\bmchi)]
	|u_0(t)\ra\la u_+(t)|{\rho}(\bmchi,t)|u_-(t)\ra\la u_0(t)|,	
\end{align}
\end{widetext}
which is expressed in the Floquet basis $|u_0(t)\ra$, $|u_\pm(t)\ra\ra$
[see \Eq{AV}].
The involved superoperators are defined as ${\cal J}[r]\rho=r\rho r^\dag$,
${\cal A}[r]\rho=r^\dag r \rho+\rho r^\dag r$, and ${\cal C}[r]\rho=[r^\dag r,\rho]$.
The first four lines describe the tunneling between QD and side reservoir
in the Lindblad-like form, with $\Gamma_{0,\pm}$ the rate for an electron
in Floquet state $|u_+(t)\ra$ or $|u_-(t)\ra$ to tunnel out of QD, and
$\Gamma_{\pm,0}$ for the opposite process.
Their connections to the tunneling rates in the original reference
$\Gam^{(\pm)}_\sgm$ ($\sgm=\upa,\dwa$) are given by \Eqs{A23} and (\ref{A24}).
The coefficients $\kappa_{0,\pm}$ and $\kappa_{\pm,0}$ arise purely from
the energy renormalization [see \Eq{A25}].

The last six lines in \Eq{ZZC} originate from the non-secular
treatment. This can be easily verified in \Eq{ZZCA}, where
all these terms are oscillating in the interaction picture.
In the case of fast oscillations, the effects of these terms
will very quickly average to zero and can thus be neglected.
\Eq{ZZC} then reduces to a Lindblad master equation, such that
the populations and coherences of the density matrix are dynamically
decoupled.
In this work, we will go beyond the secular treatment and reveal
its essential influence. The involved coefficients $\Upsilon$ and $\xi$,
defined in \Eqs{A27} and (\ref{A28}), are going to have
important roles to play in an effective magnetic field, leading
to prominent affects on the energy current.

Investigation of spin dynamics can be achieved by simply propagating
\Eq{ZZD} in the limit $\bmchi\rightarrow{\bm 0}$.
This can be done, for instance, in the Floquet basis with the diagonal
density matrix elements $\rho_{00}(t)=\la u_0(t)|\rho(t)|u_0(t)\ra$,
$\rho_{++}(t)=\la u_+(t)|\rho(t)|u_+(t)\ra$, and
$\rho_{--}(t)=\la u_-(t)|\rho(t)|u_-(t)\ra$ describing populations,
and off-diagonal density matrix elements
$\rho_{+-}(t)=\la u_+(t)|\rho(t)|u_-(t)\ra$
and
$\rho_{-+}(t)=\la u_-(t)|\rho(t)|u_+(t)\ra$ standing for coherences.
Alternatively, in this work we reexpress them in terms of
the probabilities and an average spin.
We use $\rho_{00}$ and $\rho_{11}=\rho_{++}+\rho_{--}$ to represent
the probabilities to find an empty and occupied QD, respectively.
We furthermore introduce the vector of average spin
${\bm S}=\{S_x,S_y,S_z\}$, where the individual components are given by
\be
S_x=\frac{\rho_{+-}+\rho_{-+}}{2},\,
S_y=\rmi\frac{\rho_{+-}-\rho_{-+}}{2},\,
S_z=\frac{\rho_{++}-\rho_{--}}{2}.
\ee
Therefore, the dot state is characterized by $\rho(\bmchi)=\{\rho_{00},\rho_{11},S_x,S_y,S_z\}$. According to GQME (\ref{ZZD}) and (\ref{ZZC}), it satisfies
\be\label{ZZF}
\dot{\rho}(\bmchi)={\cal L}(\bmchi)\rho(\bmchi,t),
\ee
where ${\cal L}(\bmchi)$ is a $5\times5$ matrix. Among these five equations,
two are for the occupations probabilities and three are for the average
spin.
The first two read
\begin{widetext}
\begin{align}\label{ZZBFP}
\frac{\rmd}{\rmd t}
\left(\begin{array}{c}
\rho_{00} \\
\rho_{11}
\end{array}
\right)
=&
\left(
  \begin{array}{cc}
    -\Gamma_{+,0}-\Gamma_{-,0} & \frac{1}{2}\{\Gamma_{0,+}(\bmchi)+\Gamma_{0,-}(\bmchi)\} \\
    \Gamma_{+,0}(\bmchi)+\Gamma_{-,0}(\bmchi) & -\frac{1}{2}(\Gamma_{0,+}+\Gamma_{0,-})\\
  \end{array}
\right)
\left(\begin{array}{c}
\rho_{00} \\
\rho_{11}
\end{array}
\right)
\nonumber \\
&+2
\left(\begin{array}{c}
\Upsilon_{0,+}(\bmchi)+\Upsilon_{0,-}(\bmchi) \\
-\Upsilon_{0,+}-\Upsilon_{0,-}
\end{array}
\right)S_x
+2
\left(\begin{array}{c}
\xi_{0,+}(\bmchi)-\xi_{0,-}(\bmchi) \\
-\xi_{0,+}+\xi_{0,-}
\end{array}
\right)S_y
+
\left(\begin{array}{c}
\Gamma_{0,+}(\bmchi)-\Gamma_{0,-}(\bmchi) \\
-\Gamma_{0,+}+\Gamma_{0,-}
\end{array}
\right)S_z.
\end{align}
\end{widetext}
Unambiguously, the occupation probabilities are coupled to the
spin accumulation in the QD, which is described
by the remaining three equations
\be\label{ZZBFR}
\frac{\rmd {\bm S}}{\rmd t}=\left(\frac{\rmd {\bm S}}{\rmd t}\right)_{\rm acc}
+\left(\frac{\rmd {\bm S}}{\rmd t}\right)_{\rm dec}
+\left(\frac{\rmd {\bm S}}{\rmd t}\right)_{\rm pre}.
\ee
The first term describes the spin accumulation due to tunneling between the
electrode and QD
\bea\label{ZZBFR1}
\left(\frac{\rmd {\bm S}}{\rmd t}\right)_{\rm acc}
\!\!\!\!=\!\!\!\!
&\left(\!\!
\begin{array}{cc}
\Upsilon_{-,0}(\bmchi)+\Upsilon_{+,0}(\bmchi) & -\frac{1}{2}(\Upsilon_{0,-}\!+\Upsilon_{0,+}) \\
\xi_{-,0}(\bmchi)-\xi_{+,0}(\bmchi) & -\frac{1}{2}(\xi_{-,0}-\xi_{+,0}) \\
\frac{1}{2}\{\Gamma_{+,0}(\bmchi)-\Gamma_{-,0}(\bmchi)\} & -\frac{1}{2}(\Gamma_{+,0}-\Gamma_{-,0}) \\
\end{array}
\!\!\right)
\nonumber \\
& \hspace{-5.3cm}
\cdot\left(
\begin{array}{c}
\rho_{00} \\
\rho_{11}
\end{array}
\right).
\eea
This is the source term responsible for the building up of a spin polarization
in the QD.
The second term depicts the opposite mechanism -- decay of the spin via tunneling out of spin
\be\label{ZZBFR2}
\left(\frac{\rmd {\bm S}}{\rmd t}\right)_{\rm dec}
=-\frac{1}{2}(\Gamma_{0,+}+\Gamma_{0,-}) {\bm S}.
\ee
Both spin accumulation and decay depend on the spin orientation, which is
described by the third equation
\be\label{ZZBFR3}
\left(\frac{\rmd {\bm S}}{\rmd t}\right)_{\rm pre}
={\bm S}\times \bm{{\cal B}}.
\ee
Interestingly, this term describes the precession of the average spin
about an effective magnetic field
\begin{align}\label{ZZBFU}
\bm{{\cal B}}&=\{{\cal B}_x,{\cal B}_y,{\cal B}_z\}
\nonumber \\
&=\{-(\xi_{0,+}+\xi_{0,-}),(\Upsilon_{0,+}-\Upsilon_{0,-}),\tilde{\epl}\},
\end{align}
which arises purely from the non-secular treatment and energy renormalization,
with $\tilde{\epl}=\epl_--\epl_++\frac{1}{2}(\kappa_{0,-}-\kappa_{0,+})$.
Note that this fictitious magnetic field should not be confused with
the real magnetic fields ${\bm B}$ in \Eq{Borg}.
Later, it will be shown that the subtle interaction between effective
magnetic field and spin may have an essential role to play in energy flow
through the system.

\begin{figure}
\includegraphics*[scale=0.85]{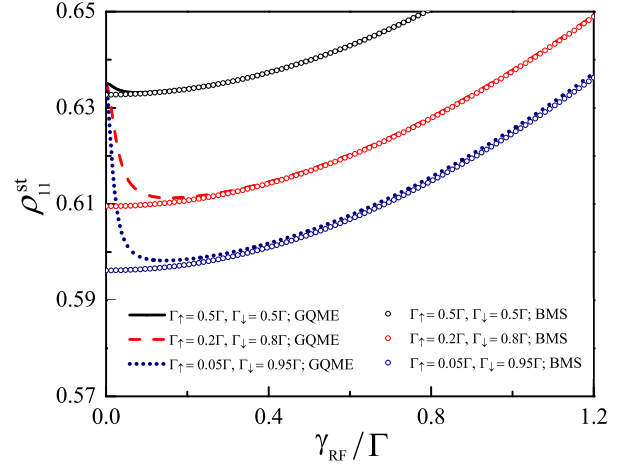}
\caption{\label{Fig3}
Comparison of stationary occupation in QD ($\rho^{\rm st}_{11}$)
using GQME (curves) and BMS master equation (symbols) for various
asymmetries in spin tunneling.
All other plotting parameters are the same as those in \Fig{Fig2}.}
\end{figure}

In literature, Bloch equation for a pseudospin has also been
derived in the singular coupling limit
(SCL) \cite{Sch09033302,Bre02,Spo80569}.
By rescaling reservoir Hamiltonian
$H_\rmB\rightarrow \lmd^{-2} H_\rmB$ and tunnel-coupling Hamiltonian
$V\rightarrow \lambda^{-1}V$, the reservoir correlation functions
experience faster decay and could even be well approximated by
$\delta(t)$ in the limit $\lmd\rightarrow0$. The SCL thus implies a flat spectral
density and high temperature limit $k_\rmB T\rightarrow\infty$, such that
the system energy splitting cannot be resolved.
The obtained pseudomagnetic fields arise purely from the energy
renormalization \cite{Sch09033302}.
The GQME employed in this work is obtained under the second order
Born-Markov approximation and thus is valid as long as the
temperature $k_\rmB T \gg \Gam $.
The preservation of positivity is guaranteed by the master equation (\ref{ZZC}),
which has also been checked numerically throughout this work.
Furthermore, the effective magnetic field originates not only from
the energy renormalization (cf. ${\cal B}_z$), but also from the
non-secular treatment (see ${\cal B}_x$ and ${\cal B}_y$), where the
latter may have even more important roles to play.

Figure \ref{Fig2} shows the stationary occupation of the QD ($\rho^{\rm st}_{11}$) and
effective spin accumulation \{$S_x^{\rm st}$, $S_y^{\rm st}$, $S_z^{\rm st}$\}
versus Rabi frequency for different configurations of asymmetry in spin tunneling.
The probability of finding an empty QD ($\rho^{\rm st}_{00}$) is not displayed as it
simply satisfies the probability conservation
$\rho^{\rm st}_{00}+\rho^{\rm st}_{11}=1$.
The occupation of the QD ($\rho^{\rm st}_{11}$) first decreases, reaches a local
minimum, and then grows rapidly towards unity with increasing Rabi frequency,
cf. \Fig{Fig2}(a).
In comparison, the results by using a Born-Markov-Secular (BMS) master equation
(i.e., by neglecting the last six lines in \Eq{ZZC}) is displayed in \Fig{Fig3}.
The results using two approaches are consistent in the regime of large Rabi
frequencies (not shown explicitly).
Yet, noticeable differences are observed for small $\gamRF$, particularly in the
case of a large asymmetry in spin tunneling.
This is due to the fact that, in the regime of small Rabi frequency, the terms
in the last six lines in \Eq{ZZCA} do not experience fast oscillation, therefore
do not average to zero.
Our results show unambiguously that it is not justified to use the secular
approximation in the regime of small Rabi frequencies.

In the limit of large Rabi frequency, an electron tunneled into the QD will be almost
localized in the Floquet state ``$|u_-(t)\ra$'' as indicated in $S^{\rm st}_z$, cf.
\Fig{Fig2}(d).
The $y$-component of the effective spin decays fast to zero, regardless of the
asymmetry in spin tunneling, as shown in \Fig{Fig2}(c).
It seems to be consistent with the usual secular treatment.
However, the $x$-component of the effective spin may survive for a wide
range of Rabi frequency, depending on asymmetry in spin tunneling, see \Fig{Fig2}(b).
We therefore emphasize that the use of a simple BMS master equation
may overlook some important dynamics of the reduced system.
We will further reveal later that the finite spin accumulation in the QD
would have a significant influence on the energy flow through the system.

\begin{figure}
\includegraphics*[scale=0.95]{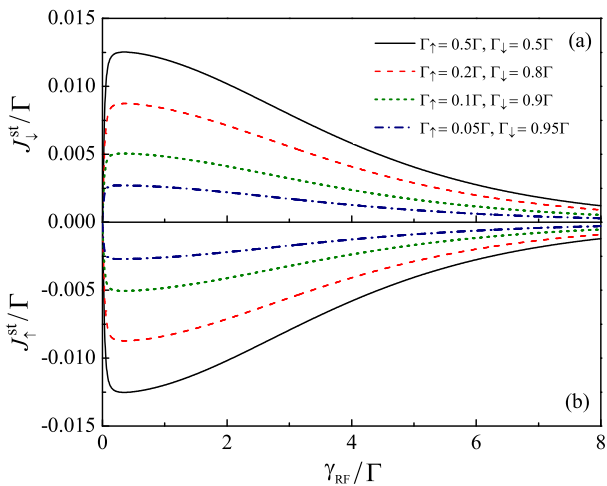}
\caption{\label{Fig4}
Individual stationary spin currents $J^{\rm st}_\dwa$ and $J^{\rm st}_\upa$ vs. Rabi frequency for various configurations of spin tunneling asymmetry.
%%
%Stationary occupation probabilities $\rho_{00}$, $\rho_{++}$, and $\rho_{--}$ %versus Rabi frequency $\gamRF$. The results obtained from BMS QME are also %plooted in dashed curves for comparison.
%%
All other plotting parameters are the same as those in  \Fig{Fig2}.}
\end{figure}

The GQME (\ref{ZZD}) and (\ref{ZZC}), or equivalently \Eq{ZZF}
enable us to evaluate various currents by employing the
CGF
\be\label{ZZE}
{\cal F}(\bmchi)=\lim_{t\rightarrow\infty}\frac{1}{t}
\ln\{\tr_\rmS[\rho(\bmchi,t)]\}
=z_0(\bmchi),
\ee
where $z_0(\bmchi)$ is the dominant eigenvalue with smallest magnitude
of ${\cal L}(\bmchi)$ defined in \Eq{ZZF} and satisfies
$z_0(\bmchi\rightarrow{\bm 0})=0$ \cite{Bag03085316}.
The individual stationary spin-$\sgm$ current can be simply obtained as
\be
J_\sgm^{\rm st}=
(-\rmi)\frac{\partial}{\partial \chi_{1\sgm}}
z_0(\bmchi)|_{\bmchi\rightarrow0}.
\ee
Throughout this work, the superscript ``st'' is used to represent
stationary values.
One straightforwardly gets the stationary spin up and spin down currents,
respectively, as
\bsube
\begin{align}
J_{\upa}^{\rm st}=&
\frac{1}{2}
(\Gamma^{(\upa)}_{0,+}+\Gamma^{(\upa)}_{0,-})
\rho^{\rm st}_{11}
-[\Gamma^{(\upa)}_{+,0}+\Gamma^{(\upa)}_{-,0}]\rho^{\rm st}_{00}
\nonumber \\
&-2[\Upsilon^{(\upa)}_{0,+}+\Upsilon^{(\upa)}_{0,-}]S^{\rm st}_x
-2[\xi^{(\upa)}_{0,+}-\xi^{(\upa)}_{0,-}]S^{\rm st}_y
\nonumber \\
&+(\Gamma^{(\upa)}_{0,+}-\Gamma^{(\upa)}_{0,-})S^{\rm st}_z,
\\
J_{\dwa}^{\rm st}=&
\frac{1}{2}
(\Gamma^{(\dwa)}_{0,+}+\Gamma^{(\dwa)}_{0,-})\rho^{\rm st}_{11}
-[\Gamma^{(\dwa)}_{+,0}+\Gamma^{(\dwa)}_{-,0}]\rho^{\rm st}_{00}
\nonumber \\
&+2[\Upsilon^{(\dwa)}_{0,+}+\Upsilon^{(\dwa)}_{0,-}]S^{\rm st}_x
+2[\xi^{(\dwa)}_{0,+}-\xi^{(\dwa)}_{0,-}]S^{\rm st}_y
\nonumber \\
&+(\Gamma^{(\dwa)}_{0,+}-\Gamma^{(\dwa)}_{0,-})S^{\rm st}_z,
\end{align}
\esube
where $\rho_{jj}^{\rm st}$ ($j=0,1$) and $S^{\rm st}_\zeta$ ($\zeta=x,y,z$) are,
respectively, the stationary solutions of Eqs. (\ref{ZZBFP}) and
(\ref{ZZBFR}) in the limit $\bmchi\rightarrow0$.

Figure \ref{Fig4} shows the individual spin currents versus Rabi frequency for various
configurations of spin tunneling asymmetry.
The spin down current ($J_\dwa^{\rm st}$) is positive as it flows out of the QD,
while the spin up current ($J_\upa^{\rm st}$) is negative as it goes into the QD.
Whenever an electron tunnels into QD, it will flow out of it. The stationary
charge current is thus zero due to charge conservation:
\be\label{ZZBFS}
J_{\rm ch}^{\rm st}=J^{\rm st}_\upa+J^{\rm st}_\dwa=0.
\ee
This is also confirmed by using \Eq{ZZBFP}.
It is worthwhile to mention that this result holds regardless of the
degree of asymmetry in spin tunneling.

\begin{figure*}
	\includegraphics*[scale=1.05]{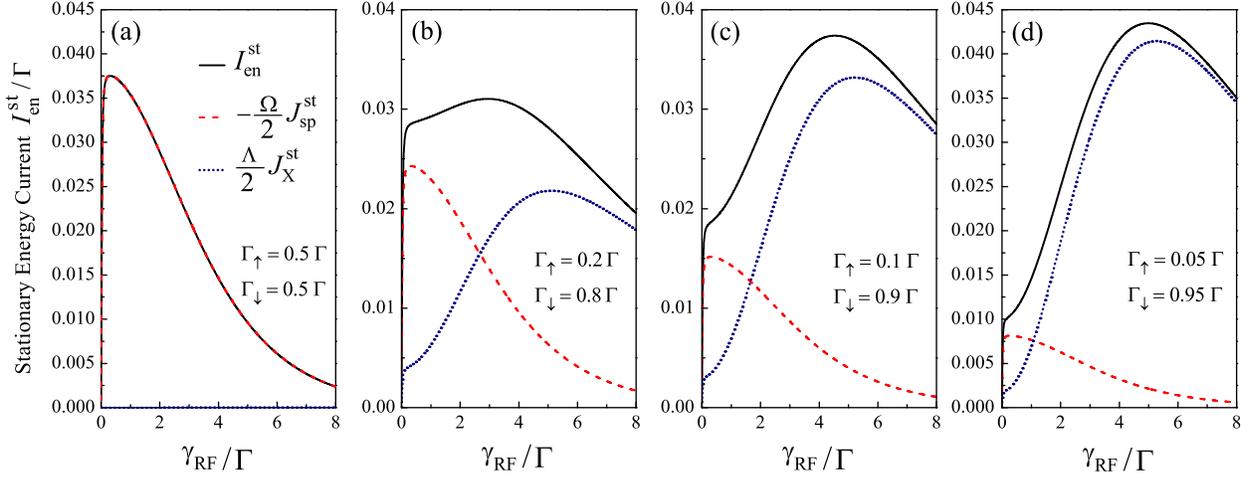}
	\caption{\label{Fig5}
		Stationary energy current vs. Rabi frequency for different configurations of
		spin tunneling asymmetry.
		For comparison, the contributions from spin current ($-\frac{\Omg}{2}J^{\rm st}_{\rm sp}$) and
		spin--magnetic field interaction ($\frac{\Lambda}{2}J^{\rm st}_X$) are also shown in dashed and dotted curves, respectively.
		All other plotting parameters are the same as those in  \Fig{Fig2}.}
\end{figure*}

In the regime of small Rabi frequency, the magnitude of either spin-up or
spin-down current increases rapidly with Rabi frequency.
%and reaches its maximum at approximately $\gamRF\simeq\Gam$.
%%
In the opposite regime of large Rabi frequency, both fall off gradually
towards zero with rising $\gamRF$.
This is due to the fact that the electron is inclined to stay in the Floquet
state ``$|u_-(t)\ra$'' as the Rabi frequency increases, cf. \Fig{Fig2}(b).
A so-called dynamical spin blockade mechanism
develops \cite{Luo15045107,Ubb13041304,Urb09165319,Wan11115304,Luo17035154},
which leads eventually to a strong suppression of the current.
An increase in spin tunneling asymmetry results in an overall inhibition
of both spin-up and spin-down currents.
With the knowledge of individual spin currents, one immediately arrives
at the net spin current, defined as
\be
J_{\rm sp}^{\rm st}\equiv J^{\rm st}_\upa-J^{\rm st}_\dwa=2J^{\rm st}_\upa,
\ee
where we have used the charge conservation, i.e. \Eq{ZZBFS}.

The stationary energy currents, associated with the spin up and spin down
currents, can be obtained in an analogous way
\be
I_\sgm^{\rm st}=(-\rmi)\frac{\partial}{\partial \chi_{2\sgm}}
z_0(\bmchi)|_{\bmchi\rightarrow0}.
\ee
By utilizing \Eq{ZZF}, one immediately arrives
at
\bsube
\begin{align}
I^{\rm st}_{\upa}=
(\epl_+-\epl_0-\textstyle{\frac{\Omg}{2}})
&\left\{\Gamma^{(\upa)}_{0,+}\rho^{\rm st}_{++}
-\Gamma^{(\upa)}_{+,0}\rho^{\rm st}_{00}\right.
\nonumber \\
&\left.-2\Upsilon^{(\upa)}_{0,+}S_x^{\rm st}
-2\xi^{(\upa)}_{0,+}S_y^{\rm st}\right\}
\nonumber \\
+(\epl_--\epl_0-\textstyle{\frac{\Omg}{2}}&)
\left\{\Gamma^{(\upa)}_{0,-}\rho^{\rm st}_{--}
-\Gamma^{(\upa)}_{-,0}\rho^{\rm st}_{00}\right.
\nonumber \\
&\left.-2\Upsilon^{(\upa)}_{0,-}S_x^{\rm st}
+2\xi^{(\upa)}_{0,-}S_y^{\rm st}\right\},
\\
I^{\rm st}_{\dwa}=
(\epl_+-\epl_0+\textstyle{\frac{\Omg}{2}})
&\left\{\Gamma^{(\dwa)}_{0,+}\rho^{\rm st}_{++}
-\Gamma^{(\dwa)}_{+,0}\rho^{\rm st}_{00}\right.
\nonumber \\
&\left.+2\Upsilon^{(\dwa)}_{0,+}S_x^{\rm st}
+2\xi^{(\dwa)}_{0,+}S_y^{\rm st}\right\}
\nonumber \\
+(\epl_--\epl_0+\textstyle{\frac{\Omg}{2}}&)
\left\{\Gamma^{(\dwa)}_{0,-}\rho^{\rm st}_{--}
-\Gamma^{(\dwa)}_{-,0}\rho^{\rm st}_{00}\right.
\nonumber \\
&\left.+2\Upsilon^{(\dwa)}_{0,-}S_x^{\rm st}
-2\xi^{(\dwa)}_{0,-}S_y^{\rm st}\right\}.
\end{align}
\esube
The total energy current in the steady state is the sum of individual
energy currents, and can be readily obtained as
\begin{align}\label{ZZBFT}
I^{\rm st}_{\rm en}=I^{\rm st}_{\upa}+I^{\rm st}_{\dwa}
=-\frac{\Omg}{2} J^{\rm st}_{\rm sp}+\frac{\Lambda}{2}J^{\rm st}_X.
\end{align}
Unambiguously, it is made up of two components.
The first contribution comes from the pure spin current $J^{\rm st}_{\rm sp}$.
The second term originates from the interaction between the accumulated
spin and the effective magnetic field
\begin{align}
J^{\rm st}_X=4(S^{\rm st}_x {\cal B}_y-S^{\rm st}_y {\cal B}_x),
\end{align}
where ${\cal B}_x$ and ${\cal B}_y$ are respectively the $x$ and $y$ components
of the effective magnetic field in \Eq{ZZBFU}.
We emphasize that the result in \Eq{ZZBFT} is of great significance in the
following two aspects.
First, an energy current can be produced even in the absence of a
net matter (charge) current. This is independent of whether the
secular approximation is made or not.
Second, the interaction between the effective magnetic field and
accumulated spin is also responsible for the production of an energy current.
This contribution purely originates from the non-secular treatment.
It will be revealed that this term may have essential roles to play in the energy
flow under the circumstance of strongly asymmetric spin tunneling rates.

In \Fig{Fig5} the stationary energy current ($I_{\rm en}^{\rm st}$) is plotted against
Rabi frequency ($\gamRF$) for various configurations of spin tunneling asymmetry.
The contributions due to spin current and spin accumulation are also exhibited
by the dashed and dotted curves, respectively.
In the case of symmetric spin tunneling ($\Gam_\upa=\Gam_\dwa=0.5\Gam$),
the spin accumulation ($J^{\rm st}_X$) has a negligible contribution and the
energy current is dominated by the spin current ($J^{\rm st}_{\rm sp}$),
cf. \Fig{Fig5}(a).
The behavior of the total energy current is thus similar to the spin current in
\Fig{Fig4}, i.e., it first increases rapidly with rising Rabi frequency, reaches a
maximum at approximately $\gamRF \approx 0.5\Gam$,
and finally falls off and approaches zero in the limit $\gamRF\rightarrow\infty$.

The picture becomes drastically different under the circumstance
of strong asymmetry in spin tunneling, where the spin current is
suppressed while the contribution from interaction between effective
magnetic field and spin accumulation has an essential role to play.
For instance, a strong asymmetry of $\Gam_\upa=0.2\Gam$ and $\Gam_\dwa=0.8\Gam$
gives rise to a prominent enhancement in $\frac{\Lmd}{2}J^{\rm st}_X$
at approximately $\gamRF\approx 5\Gam$.
This is ascribed to noticeable effective magnetic field ${\cal B}_y$ [inset of \Fig{Fig2}(b)]
and finite spin accumulation $S^{\rm st}_x$ [\Fig{Fig2}(b)] ($S^{\rm st}_y$ has a
vanishing contribution, cf. \Fig{Fig2}(c)).
It leads to the emergence of a local maximum in
the energy current at $\gamRF\approx 3\Gam$, in addition to
its original maximum at $\gamRF\approx 0.5\Gam$, see \Fig{Fig5}(b).
Remarkably, for an extremely asymmetric
tunneling rates ($\Gam_\upa=0.05\Gam,\Gam_\dwa=0.95\Gam$) as shown in \Fig{Fig5}(d),
the $J^{\rm st}_X$-term serves as the dominant contribution.
As a result, one observes solely a single maximum in energy current
at $\gamRF\approx 5\Gam$.
Our results show unambiguously the importance of considering the interaction between
effective magnetic field and spin accumulation for very asymmetric spin tunneling rates.
In this case, the use of a simple BMS master equation picture would be
fundamentally insufficient.

\section{\label{thsec4}Stochastic Thermodynamics}

Now we are in a position to discuss the stochastic thermodynamics, i.e.,
the identification of the first and second laws at the microscopic level.
Our analysis is based on the two-measurement process theory and the method
of counting statistics \cite{Esp091665}.
We will focus on energy and entropy balance of the ESR pumped QD
system in the steady-state limit.

\subsection{Energy balance based on statistics of \\ work and heat}

In this subsection we characterize the steady fluctuations of
work and heat by evaluating their counting statistics
using initial and final measurements of the system energy.
Statistics of work has previously been analyzed with quantum jump approach \cite{Hek13093602}
or Lindblad quantum master equations \cite{Sil14022103}. Here we employ the
two-measurement process approach \cite{Esp091665,Sol13060508}
within the framework of the GQME, thus fully accounting
for effect of the non-secular treatment and the coherences.

Let us denote the instantaneous eigenenergies of $H_\rmS(t)$ as $e_m(t)$
and those of $H_\rmB$ as $\vpl_k$. Assume that at time $t=0$ a joint
measurement of $H_\rmS(0)$ and $H_\rmB$  is performed, yielding the
outcomes $e_m(0)$ and $\vpl_k$, respectively.
At time $t>0$, a second joint measurement of $H_\rmS(t)$ and $H_\rmB$ is
made with outcomes $e_n(t)$ and $\vpl_{k'}$.
The energy changes of system and bath, $\Dlt e$ and $\Dlt \vpl$,
respectively, in a single realization of the protocol are thus given by
\bsube
\begin{align}
\Dlt e&=e_n(t)-e_m(0),
\\
\Dlt \vpl &=\vpl_{k'}-\vpl_k.
\end{align}
\esube
The change of the energy of the entire system is given by the sum of the energy
changes of the reduced system and bath energies, apart from a negligible
contribution due to the system-bath interaction.
The work $W$ performed on the system coincides with the change of the
energy of the entire system because the driving is acted solely onto the reduced system,
i.e.,
\be
W=\Dlt e+\Dlt \vpl=e_n(t)+\vpl_{k'}-e_m(0)-\vpl_k.
\ee
Apparently, $W$ is a random variable due to the intrinsic randomness in
the quantum measurement processes.
The statistical properties of $W$ can be conveniently expressed in
terms of its characteristic function
\be\label{ZZA}
G_W(\chi_W,t)=\int_{-\infty}^\infty \rmd W e^{\rmi W \chi_W} P(W,t),
\ee
where $P(W,t)$ is probability density function of observing an amount
of work $W$ performed by the external driving from time 0 to $t$, and
$\chi_W$ is the corresponding counting field.
According to the theory of two-measurement process \cite{Esp091665},
the probability density function is given by
\begin{align}\label{ZZB}
P(W,t)\!=&\!\!\sum_{e_n(t),e_m(0)}\sum_{\vpl_k,\vpl_{k'}}
\!\dlt[e_n(t)\!+\vpl_{k'}\!-e_m(0)-\vpl_k-W]
\nonumber \\
&\cdot p(e_n(t)\!+\vpl_{k'}|e_m(0)\!+\vpl_k)p(e_m(0)\!+\vpl_k),
\end{align}
where $p(e_n(t)+\vpl_{k'}|e_m(0)+\vpl_k)$ is the conditional probability that a measurement of $H_\rmS(t)$ and $H_\rmB$ gives $e_n(t)$ and $\vpl_{k'}$, respectively, at time $t$ given that it gave $e_m(0)$ and $\vpl_k $ at time 0, while $p(e_m(0)+\vpl_k)$ is the usual probability to have $e_m(0)$ and $\vpl_k $ at time 0.

By introducing the projectors $\hat{\cal P}_{e_j(t)}$ and $\hat{\cal P}_{\vpl_k}$
on the $j$-th state of the system with energy $e_j(t)$ and $k$-th state of the reservoir with energy $\vpl_k$, one has
\begin{align}\label{ZZ5}
&p(e_n(t)+\vpl_{k'}|e_m(0)+\vpl_k)p(e_m(0)+\vpl_k)
\nonumber \\
&=
\tr[\hat{\cal P}_{e_n(t)}\hat{\cal P}_{\vpl_{k'}} U(t)
\hat{\cal P}_{e_m(0)}\hat{\cal P}_{\vpl_k}
\rho_\rmT(0)
\nonumber \\
&\hspace{5ex}\times\hat{\cal P}_{\vpl_k}\hat{\cal P}_{e_m(0)}
U^\dag(t)\hat{\cal P}_{\vpl_{k'}} \hat{\cal P}_{e_n(t)}],
\end{align}
where $\tr[\cdots]$ represents the trace over the degrees of the
freedom of entire system, and  $U(t)$ is the evolution operator
associated with the entire Hamiltonian $H_\rmT(t)$:
\be
U(t)=\exp_+\left(-\rmi \int_0^t \rmd \tau H_\rmT (\tau)\right).
\ee
Next we assume that the initial total density matrix can be factorized to
$\rho_{\rm T}(0)=\frac{e^{-\beta H_\rmS(0)}}{Z_\rmS(0)}\otimes
\frac{e^{-\beta H_\rmB}}{Z_\rmB}$, where
$Z_\rmS(0)=\tr_\rmS\{e^{-\beta H_\rmS(0)}\}$ is the
partition function of system at time $t=0$ and
$Z_\rmB=\tr_\rmB\{e^{-\beta H_\rmB}\}$ is that of reservoir.
By using
$\hat{\cal P}^2_{e_j(t)}=\hat{\cal P}_{e_j(t)}$ and
$\hat{\cal P}^2_{\vpl_k}=\hat{\cal P}_{\vpl_k}$, \Eq{ZZ5} is further
simplified to
\begin{align}\label{ZZ6}
&p(e_n(t)+\vpl_{k'}|e_m(0)+\vpl_k)p(e_m(0)+\vpl_k)
\nonumber \\
&=\tr[U^\dag(t) \hat{\cal P}_{e_n(t)}\hat{\cal P}_{\vpl_{k'}} U(t)
\hat{\cal P}_{e_m(0)}\hat{\cal P}_{\vpl_k}\rho_\rmT(0)].
\end{align}
Noticing that
$\sum_{e_j(t)} \hat{{\cal P}}_{e_j(t)}e^{\pm\rmi\chi_W e_j(t)}
=e^{\pm\rmi \chi_W H_\rmS(t)}$
and
$\sum_{\vpl_k} \hat{{\cal P}}_{\vpl_k}e^{\pm\rmi\chi_W \vpl_k}
=e^{\pm\rmi\chi_W H_\rmB}$, the characteristic function of work in \Eq{ZZA} can be written as
\begin{align}
{G}_W(\chi_W,\!t)\!=&\tr\!\left\{\!e^{\rmi\frac{\!\chi_W}{2} [H_\rmS(t)+H_\rmB]}
U(t)e^{-\rmi\frac{\!\chi_W}{2}[H_\rmS(0)+H_\rmB]}\! \rho_{\rm T}(0) \right.
\nonumber \\
&\;\;\; \cdot\left.
e^{-\rmi\frac{\chi_W}{2} [H_\rmS(0)+H_\rmB]}
U^\dag(t) e^{\rmi\frac{\chi_W}{2}[H_\rmS(t)+H_\rmB]}
\right\},
\nonumber \\
=&\sum_{e_n(t),e_m(0)} \frac{e^{\rmi \chi_W [e_n(t)-e_m(0)]-\beta e_m(0)}}{Z_\rmS(0)}
\nonumber \\
&\;\;\; \cdot\la e_n(t)|\varrho(\chi_W,t;e_m(0))|e_n(t) \ra,
\end{align}
where we have introduced a new density matrix of the reduced system
including the counting field of work
\begin{align}\label{varrhow}
\varrho(\chi_W,t;e_m(0))=&\tr_\rmB \left\{e^{\rmi\frac{\chi_W}{2} H_\rmB}
U(t)e^{-\rmi\frac{\chi_W}{2} H_\rmB} \varrho(0;e_m(0)) \right.
\nonumber \\
&\hspace{3ex}\left.\otimes\,\rho_\rmB\,
e^{-\rmi\frac{\chi_W}{2} H_\rmB}
U^\dag(t) e^{\rmi\frac{\chi_W}{2}H_\rmB}\right\}.
\end{align}
By comparing with \Eqs{rhoTchi} and (\ref{Hpchi}), one finds that
$\varrho(\chi_W,t;e_m(0))$ satisfies the same equation as $\rho(\bmchi,t)$
does, i.e. \Eq{ZZF}, with only the crucial replacement
\be
\varrho(\chi_W,t;e_m(0))=\rho(\bmchi,t)|_{\chi_{1\upa}=\chi_{1\dwa}=0,\chi_{2\upa}=\chi_{2\dwa}=\chi_W}
\ee
and initial condition $\varrho(0;e_m(0))=|e_m(0)\ra \la e_m(0)|$.
Unambiguously, both populations and coherences have important
roles to play in the statistics of work.
Here, we are interested in the stationary statistics, therefore, the CGF
of the mechanical power is simply given by
\begin{align}
{\cal G}_W(\chi_W,t)=&\lim_{t\rightarrow\infty} \frac{1}{t}\ln G_W (\chi_W,t)
\nonumber \\
=&z_0(\bmchi)|_{\chi_{1\upa}=\chi_{1\dwa}=0,
	\chi_{2\upa}=\chi_{2\dwa}=\chi_W},
\end{align}
where $z_0(\bmchi)$ is the dominant eigenvalue with smallest magnitude
of ${\cal L}(\bmchi)$ as shown in \Eq{ZZE}.

The average rate of mechanical work is simply obtained
\begin{align}\label{ZZH}
\dot{W}&=-\rmi \frac{\partial }{\partial \chi_W}
z_0(\{0,0,\chi_W,\chi_W\})|_{\chi_W\rightarrow0}
\nonumber \\
&=I_{\upa}^{\rm st}+I_{\dwa}^{\rm st}
=-\frac{\Omg}{2} J^{\rm st}_{\rm sp}+\frac{\Lambda}{2}J^{\rm st}_X.
\end{align}
It shows clearly that the work done on the system is used to
produce a spin current and a precession of an average spin in the
QD, where the latter originates purely from non-secular treatment.

The statistical properties of the heat flow can be analyzed in a similar way
by using the two-measurement process theory \cite{Esp091665,Gas14115001}.
Assume at time $t=0$ a measurement $H_\rmB$ yields an outcome $\vpl_k$.
A little later at $t>0$, a second measurement is made with outcome
$\vpl_{k'}$. The heat flows from the reservoir to the QD is given by
\be
Q=-(\vpl_{k'}-\vpl_{k}).
\ee
Analogously to \Eq{ZZA}, the characteristic function of heat flow
is defined as
\be
G_Q(\chi_Q,t)=\int_{-\infty}^\infty \rmd Q e^{\rmi Q \chi_Q} P(Q,t),
\ee
where $\chi_Q$ is the counting field associated with heat flow, and
$P(Q,t)$ is the probability density function of observing an amount
of heat $Q$ flowed in to QD from time 0 to $t$. It can be determined
from the two-time measurement approach \cite{Esp091665}:
\be
P(Q,t)=\sum_{\vpl_k,\vpl_{k'}}\dlt[\vpl_{k'}-\vpl_k+Q)]
p(\vpl_{k'}|\vpl_k)p(\vpl_k),
\ee
where $p(\vpl_{k'}|\vpl_k)$ is the conditional probability for observing $\vpl_{k'}$ at time $t$, given that it yields $\vpl_k $ at time 0, while $p(\vpl_k)$ is the usual probability to have $\vpl_k $ at time 0. By
following similar procedures in \Eqs{ZZ5} and (\ref{ZZ6}), the characteristic function of heat flow can be expressed as
\begin{align}
{G}_Q(\chi_Q,t)=\tr_\rmS[\varrho(-\chi_Q,t)],
\end{align}
where $\varrho(-\chi_Q,t)$ satisfies the same equation as $\rho(\bmchi,t)$
in \Eq{ZZF}, with only the replacement $\varrho(-\chi_Q,t)=
\rho(\bmchi,t)|_{\chi_{1\upa}=\chi_{1\dwa}=0,\chi_{2\upa}=\chi_{2\dwa}=-\chi_Q}$.
In comparison with the \Eq{ZZE}, one immediately finds the CGF of the heat
flow in the stationary limit
\begin{align}
{\cal G}_Q(\chi_Q)&=\lim_{t\rightarrow\infty}
\frac{1}{t}\ln G_Q (\chi_Q,t)
\nonumber \\
&=z_0(\bmchi)_{\chi_{1\upa}=\chi_{1\dwa}=0,
	\chi_{2\upa}=\chi_{2\dwa}=-\chi_Q},
\end{align}
where $z_0(\bmchi)$ is the dominant eigenvalue with smallest
magnitude of ${\cal L}(\bmchi)$ in \Eq{ZZE}.
The average heat flowing into QD is simply given by
\begin{align}\label{ZZK}
\dot{Q}&=-\rmi \frac{\partial }{\partial \chi_Q}
z_0(\{0,0,-\chi_Q,-\chi_Q\})|_{\chi_Q\rightarrow0}
\nonumber \\
& =-(I_{\upa}^{\rm st}+I_{\dwa}^{\rm st}).
\end{align}
By comparing with \Eqs{ZZH}, one eventually arrives at the energy balance
relation in terms of the first law of thermodynamics
\be\label{dotE}
\dot{E}=\dot{Q}+\dot{W}=0.
\ee
The work done on the system is completely converted into heat, such that
the net increase of energy in the QD is zero in the stationary limit.

\subsection{Entropy balance}

We are now in a position to investigate the entropy balance and
identify the second law on the microscopic level.
For a system described by a Lindblad master equation, it has
been shown that a connection
between entropy production and the heat currents can be established
based on the local detailed balance (LDB) relation \cite{Esp10011143}.
A similar LDB was found in a driven system under secular
approximation, where the populations and coherences are dynamically
decoupled \cite{Sil14022103,Cue15055002}.
Our analysis is based on the GQME beyond the secular
approximation such that both populations and coherences have
vital roles to play.
We will establish the connection between entropy production and heat
flow in the stationary limit, using the two-measurement process theory
and the method of counting statistics, rather than the LDB.

We start with the von Neumann entropy
\be
S(t) =-\tr_\rmS [\rho(t) \ln \rho(t)],
\ee
where $\rho(t)$ is the density matrix of the reduced system.
The change in von Neumann entropy $\Dlt S$ can be decomposed into an entropy
production $\Dlt S_\rmi$ and an entropy
flow $\Dlt S_{\rm e}$ \cite{Kon98,Gro84,Pri77,Gas06201},
where the latter is given by the heat exchanged with the
reservoir multiplied by the inverse temperature
\be\label{ZZU}
\Dlt S_{\rm e}=\beta \Dlt Q.
\ee
The entropy production then is given by
\be\label{ZQa}
\Dlt S_\rmi= \Dlt S-\beta \Dlt Q.
\ee
In what follows, we will analyze the entropy from a statistical point of
view, using the two-measurement process analogous to that in \Eq{ZZB}.
Assume at time $t=0$, joint measurements of the energies of system and
reservoir are made. The obtained energy eigenvalues are denoted
by $e_m(0)$ and $\epl_k$, respectively.
The associated von Neumann entropy of the system is
${S}(0)=\beta [e_{m}(0)-{F}_0]$, where ${F}_0=-\frac{1}{\beta}\ln {Z}_0$ is
the free energy with $\ln {Z}_0=-\ln p(e_m(0))-\beta e_m(0)$.
Later at time $t$, a similar joint measurement yields $e_n(t)$ and $\epl_{k'}$,
respectively.
The corresponding entropy thus is ${S}(t)=\beta [e_n(t)-{F}_t]$,
with free energy ${F}_t=-\frac{1}{\beta}\ln {Z}_t$ and
$\ln{Z}_t=-\ln p(e_n(t))-\beta e_n(t)$.
The change in entropy is
\be
\Dlt S=S(t)-S(0)=\ln p(e_m(0))-\ln p(e_n(t)).
\ee
Apparently, $\Dlt S$ is a random variable owing to the stochastic nature in
the quantum measurement. Its probability density function is given by
\begin{align}
P(\Dlt S,t)=
&\!\!\!\!\sum_{e_n(t),e_m(0)}\sum_{\vpl_k,\vpl_{k'}}
\!\!\dlt[\ln p(e_m(0))\!-\ln p(e_n(t))\!-\!\Dlt S]
\nonumber \\
&\cdot p(e_n(t)+\vpl_{k'}|e_m(0)+\vpl_k)p(e_m(0)+\vpl_k),
\end{align}
where we have used the same probabilities $p(e_n(t)+\vpl_{k'}|e_m(0)+\vpl_k)$
and $p(e_m(0)+\vpl_k)$ as in \Eq{ZZB}.
In a single realization the entropy can be expressed as
\be
\Dlt S=\Dlt S_\rmi + \beta \Dlt Q,
\ee
where $\Dlt Q=-(\vpl_{k'}-\vpl_k)$  is the heat flowed into the QD.
One finds the probability density function for entropy production:	
\begin{align}
P(\Dlt S_\rmi,t)=
&\sum_{e_n(t),e_m(0)}\sum_{\vpl_k,\vpl_{k'}}
\dlt[\ln p(e_m(0))-\ln p(e_n(t))
\nonumber \\
&\hspace{2.5cm}-\Dlt S_\rmi+\beta (\vpl_{k'}-\vpl_k)]
\nonumber \\
&\cdot p(e_n(t)+\vpl_{k'}|e_m(0)+\vpl_k) p(e_m(0)+\vpl_k).
\end{align}
Its corresponding characteristic function can be obtained by
simply performing a Fourier transform
\begin{align}
G_{\Dlt S_\rmi}(\chi_{S_i})&=\int_{-\infty}^{\infty} \rmd (\Dlt S_\rmi)
P(\Dlt S_\rmi,t) \, e^{\rmi \Dlt S_\rmi \chi_{S_i} },
\end{align}
where $\chi_{S_i}$ is the counting field associated with the
entropy production.
In parallel to the discussion for the statistics of work, it
can be expressed as
\begin{align}
G_{\Dlt S_\rmi}(\chi_{S_i},t)=
&\!\!\!\sum_{e_n(t),e_m(0)}\!\!\!
\frac{e^{\rmi \chi_{S_i} [\ln p(e_m(0))-\ln p(e_n(t))]-\beta e_m(0)}}{Z_\rmS(0)}
\nonumber \\
&\;\;\;\;\;\;\cdot \la e_n(t)|\varrho(\chi_{S_i} \beta,t; e_m(0))|e_n(t)\ra,
\end{align}
where $\varrho(\chi_{S_i} \beta,t; e_m(0))$
satisfies the same equation as $\rho(\bmchi,t)$ in \Eq{ZZF}, with only the
replacement $\varrho(\chi_{S_i}\beta,t;e_m(0))=\rho(\bmchi,t)|_{\chi_{1\upa}
=\chi_{1\dwa}=0,\chi_{2\upa}=\chi_{2\dwa}=\chi_{S_i}\beta}$
and initial condition $\varrho(0;e_m(0))=|e_m(0)\ra \la e_m(0)|$.

The CGF for the entropy production characterizing the stationary statistics
is thus given by
\begin{align}
{\cal G}_{S_\rmi}(\chi_{S_i})
=&\lim_{t\rightarrow\infty}
\frac{1}{t}\ln G_{\Dlt S_\rmi} (\chi_{S_i},t)
\nonumber \\
=&z_0(\bmchi)|_{\chi_{1\upa}
=\chi_{1\dwa}=0,\chi_{2\upa}=\chi_{2\dwa}=\chi_{S_i}\beta},
\end{align}
where $z_0(\bmchi)$ is the dominant eigenvalue with smallest magnitude
of ${\cal L}(\bmchi)$, as shown in \Eq{ZZE}.
The average irreversible entropy production then is simply obtained
by taking partial derivative of ${\cal G}_{S_\rmi}(\chi_{S_i})$
\begin{align}\label{ZZK}
\dot{S_\rmi}&=
-\rmi \frac{\partial }{\partial \chi_{S_i}}
z_0(\{0,0,\chi_{S_i}\beta,\chi_{S_i}\beta\})|_{\chi_{S_i}\rightarrow0}
\nonumber \\
&=\beta (I_{\upa}^{\rm st}+I_{\dwa}^{\rm st})
=\beta I_{\rm en}^{\rm st} \ge 0.
\end{align}
This is the second law formulated as the non-negativity of
the irreversible entropy production, owing to the non-negative
energy current that we have confirmed numerically, see also \Fig{Fig5}.
Furthermore, it means that both populations and coherences of the
density matrix have essential roles to play in entropy production.
By recalling \Eq{ZZH} and the energy balance relation \Eq{dotE},
one finds
\be\label{dotSi}
\dot{S}_\rmi =-\beta \dot{Q}=\frac{\beta}{2}(-\Omega J_{\rm sp}^{\rm st}
+\Lambda J_X^{\rm st}).
\ee
The first equality manifests the relation between entropy
production and heat current based on counting statistics, rather than the LDB.
Together with the entropy flow defined in \Eq{ZZU}, one
arrives at the entropy balance in the stationary limit
\be
\dot{S}=\dot{S}_\rmi+\dot{S}_{\rm e}=0.
\ee
The second equality in \Eq{dotSi}, to the best of our knowledge, is a new
result, revealing not only its relation to a pure spin current, but also the intimate
connection to the influence of interaction between effective spin and effective magnetic field.
Finally, we remark that the contribution from $J_X^{\rm st}$ is of the first
order in tunnel-coupling strength, and thus could be detected in experiment.
We highly anticipate this to be verified in the near future.

\section{\label{thsec5}Conclusion}

In summary, we have performed a stochastic thermodynamics analysis of
an ESR pumped quantum dot system in the presence of a pure spin current
only.
The state of the system can be described by populations and an effective
spin in the Floquet basis. In particular, this effective spin undergoes a
precession about an effective magnetic field, which originates from the
non-secular treatment and energy renormalization.
Unambiguously, an energy current could be generated, not due to a charge current,
but entailed by a pure spin current and the interaction between effective
spin and effective magnetic field, where the latter may have the dominant
role to play in the case of strong asymmetry in spin tunneling.
In the stationary limit, energy balance and entropy balance relations are
established based on the theory of counting statistics.
Furthermore, we revealed a new mechanism that the irreversible
entropy production is found to be intimately related to the interaction
between effective spin and magnetic field.

 \begin{acknowledgments}
 We would like to thank S. K. Wang for the fruitful discussion.
 Support from the National Natural Science Foundation of China
 (Grant Nos. 11774311 and 11647082) and education department of
 Zhejiang Province (No. 8) is gratefully acknowledged.
 \end{acknowledgments}

\appendix

\section{\label{App-A} Derivation of the GQME}

First, we transform from the Schr\"{o}dinger's picture to the
interaction picture
\be
\tilde{\rho}_\rmT(\bmchi,t)=U^\dag_0(t)\rho_\rmT(\bmchi,t) U_0(t),
\ee
where
\begin{gather}\label{AS}
U_0(t)\equiv \exp_+\left\{-\rmi \int_0^t \rmd \tau H_\rmS(\tau)\right\}
e^{-\rmi H_\rmB t}.
\end{gather}
In what follows, the tilde is used to indicate a quantity in the interaction picture.
The equation of motion of $\tilde{\rho}_{\rm T}(\bmchi,t)$ then reads
\be\label{A5}
\frac{\rmd}{\rmd t}{\tilde{\rho}}_\rmT(\bmchi,t)=
-\rmi\{\tilde{V}(\bmchi,t)\tilde{\rho}_\rmT(\bmchi,t)-
\tilde{\rho}_\rmT(\bmchi,t)\tilde{V}(-\bmchi,t)\},
\ee
with
\be\label{A5X}
\tilde{V}(\bmchi,t)=U_0^\dag(t) V(\bmchi) U_0(t).
\ee

Now, we integrate \Eq{A5} twice, differentiate with
respect to time ``$t$'', and trace over the degrees of
freedom of the reservoir.
This yields an exact equation of motion for the $\bmchi$-dependent
reduced density matrix
\begin{align}\label{A6X}
\frac{\rmd}{\rmd t}{\tilde{\rho}}(\bmchi,t)=-\int_0^t
&\rmd \tau \,
\tr_{\rm B}\{\tilde{V}(\bmchi,t)\tilde{V}(\bmchi,\tau)\tilde{\rho}_\rmT(\bmchi,\tau)
\nonumber \\
&-\tilde{V}(\bmchi,\tau)\tilde{\rho}_\rmT(\bmchi,\tau)\tilde{V}(-\bmchi,t)
\nonumber \\
&-\tilde{V}(\bmchi,t)\tilde{\rho}_\rmT(\bmchi,\tau)\tilde{V}(-\bmchi,\tau)
\nonumber \\
&+\tilde{\rho}_\rmT(\bmchi,\tau)\tilde{V}(-\bmchi,\tau)\tilde{V}(-\bmchi,t)\},
\end{align}
where $\tr_\rmB\{\cdots\}$ stands for the trace over the degrees of freedom of reservoir.
This equation still contains the density matrix $\tilde{\rho}_{\rm T}(\bmchi,\tau)$
of the entire system.
We thus now make the Born approximation which assumes that the density
operator factorises at all times as
$\tilde{\rho}_\rmT(\bmchi,\tau)=\tilde{\rho}(\bmchi,\tau)\otimes\rho_{\rm B}$.
It greatly simplifies \Eq{A6X}. Yet, it still has a time non-local form: The future evolution
${\tilde{\rho}}(\bmchi,t)$ depends on its past history ${\tilde{\rho}}(\bmchi,\tau)$, which makes it difficult to work with.
In case of a large separation between system and environment
timescales, it is justified to introduce the Markov approximation,
i.e., replacing $\tilde{\rho}(\bmchi,\tau)$ by $\tilde{\rho}(\bmchi,t)$ and
extending the upper limit of the integral to infinity in \Eq{A6X}.
Finally, it yields a closed differential equation of motion for the reduced
density matrix that the future behaviour of ${\tilde{\rho}}(\bmchi,t)$
depends only on its present state
\begin{align}\label{B6}
\frac{\rmd}{\rmd t}{\tilde{\rho}}(\bmchi,t)
=&-\!\int_{0}^\infty
\!\rmd \tau
\tr_{\rm B}\{\tilde{V}(\bmchi,t)\tilde{V}(\bmchi,t-\tau)\rho_\rmB\!\otimes
\tilde{\rho}(\bmchi,t)
\nonumber \\
&-\tilde{V}(\bmchi,t-\tau)\rho_{\rm B}\!\otimes \tilde{\rho}(\bmchi,t)\tilde{V}(-\bmchi,t)
\nonumber \\
&-\tilde{V}(\bmchi,t)\rho_{\rm B}\!\otimes \tilde{\rho}(\bmchi,t)\tilde{V}(-\bmchi,t-\tau)
\nonumber \\
&+\rho_{\rmB}\!\otimes\tilde{\rho}(\bmchi,t)\tilde{V}(-\bmchi,t-\tau)\tilde{V}(-\bmchi,t)\}
\nonumber \\
=&-\{{\rm [I]}-{\rm [II]}-{\rm [III]}+{\rm [IV]}\}.
\end{align}
This serves as an essential starting point for the following
derivation.

Each term in \Eq{B6} has to be evaluated, which requires the explicit
form of the $V(\bmchi,t)$ in \Eq{A5X}.
By utilizing \Eqs{ARB} and (\ref{AS}), one immediately has
\begin{align}\label{AW}
\tilde{f}_{\sgm}(\bmchi,t)=
& U^\dag_0(t) {f}_{\sgm}(\bmchi) U_0(t)
\nonumber \\
=&\sum_{k}t_{k\sgm}c_{k\sgm}
e^{-\frac{\rmi}{2}(\chi_{2\sgm}\vpl_{k\sgm}+\chi_{1\sgm})}
e^{-\rmi\vpl_{k\sgm}t}.
\end{align}
The system operators in the interaction picture are defined
analogously
\begin{align}\label{AT}
\tilde{d}_\sgm(t)&=U^\dag_0(t) {d}_\sgm(t) U_0(t)
\nonumber \\
&=\exp_- \left\{\rmi\int_0^t H_\rmS(\tau) d\tau\right\}
d_\sgm \exp_+\left\{-\rmi\int_0^t H_\rmS(\tau) d\tau\right\},
\end{align}
where the time ordering arises purely from the time dependence of the
system Hamiltonian.
To explicitly evaluate \Eq{AT}, we employ
Floquet theory \cite{Shi65B979,Hol941950,Gri98229,Szc13012120},
which states that the unitary evolution can be represented as
\be\label{AU}
\exp_+\left\{\!-\rmi\int_0^t H_\rmS(\tau) d\tau\right\}=\sum_j e^{-\rmi \epl_j t}|u_j(t)\ra\la u_j(0)|,
\ee
where $|u_j(t)\ra$ is the Floquet function inherits the periodicity
$|u_j(t)\ra=|u_j(t+T)\ra$ with $T=\frac{2\pi}{\Omg}$,
and $\epl_j$ is the corresponding quasienergy.
The quasienergies and Floquet functions are simply
given, respectively, by
\bsube\label{AV}
\begin{align}
\epl_0&=0, \hspace{1.1cm} |u_0(t)\ra=
\left(\begin{array}{c}
	1 \\
	0 \\
	0
\end{array}\right),
\\
\epl_+&=\frac{\Omg+\Lambda}{2},\;\;\;
|u_+(t)\ra=\left(\begin{array}{c}
0 \\
+\sin(\frac{\Theta}{2})e^{+\rmi\hfOmg t} \\
+\cos(\frac{\Theta}{2})e^{-\rmi\hfOmg t}
\end{array}\!\right),
\\
\epl_-&=\frac{\Omg-\Lambda}{2},\;\;\;
|u_-(t)\ra=\left(\begin{array}{c}
0 \\
-\cos(\frac{\Theta}{2})e^{+\rmi\hfOmg t} \\
+\sin(\frac{\Theta}{2})e^{-\rmi\hfOmg t}
\end{array}\!\right),
\end{align}
\esube
where, for brevity, we have introduced
\bsube
\begin{align}
\sin\left(\frac{\Theta}{2}\right)&=\sqrt{\frac{\Lambda+\dlt}{2\Lambda}},
\\
\cos\left(\frac{\Theta}{2}\right)&=\sqrt{\frac{\Lambda-\dlt}{2\Lambda}},
\end{align}
\esube
with $\dlt=\Dlt-\Omega$ is the ESR detuning
and $\Lambda=\sqrt{\dlt^2+4\gamRF^2}$.
The annihilation operators of the system in the Floquet basis can thus be
readily expressed as
\bsube\label{ZZ1}
\begin{align}
\tilde{d}_\upa(t)=&
\sin ({\textstyle \frac{\Theta}{2}})e^{-\rmi(\epl_+-\epl_0-\hfOmg)t}
|u_0(0)\ra\la u_+(0)|
\nonumber \\
&-\cos ({\textstyle \frac{\Theta}{2}})e^{-\rmi(\epl_--\epl_0-\hfOmg)t}
|u_0(0)\ra\la u_-(0)|,
\\
\tilde{d}_\dwa(t)=&
\cos ({\textstyle \frac{\Theta}{2}})e^{-\rmi(\epl_+-\epl_0+\hfOmg)t}
|u_0(0)\ra\la u_+(0)|
\nonumber \\
&+\sin ({\textstyle \frac{\Theta}{2}})e^{-\rmi(\epl_--\epl_0+\hfOmg)t}
|u_0(0)\ra\la u_-(0)|.
\end{align}
\esube
Their corresponding creation operators can be obtained by simply taking the
Hermitian conjugate.

The following procedure relies on the substituting of
\Eqs{AW} and (\ref{ZZ1}) into \Eq{B6}.
For instance, the first term [I] in \Eq{B6} is obtained as
\begin{widetext}
\begin{align}\label{ZZ1A}
{\rm [I]}=
&\left\{\gam^{(+)}_{\dwa}(\epl_+-\epl_0+\hfOmg)\cos^2 ({\textstyle \frac{\Theta}{2}})
+\gam^{(+)}_{\upa}(\epl_+-\epl_0-\hfOmg)\sin^2 ({\textstyle \frac{\Theta}{2}})\right\}
|u_0(0)\ra\,\la u_0(0)|\,\tilde{\rho}(\bmchi,t)
\nonumber \\
&+
\left\{\gam^{(+)}_{\dwa}(\epl_--\epl_0+\hfOmg)\sin^2 ({\textstyle \frac{\Theta}{2}})
+\gam^{(+)}_{\upa}(\epl_--\epl_0-\hfOmg)\cos^2 ({\textstyle \frac{\Theta}{2}})\right\}
|u_0(0)\ra\,\la u_0(0)|\,\tilde{\rho}(\bmchi,t)
\nonumber \\
&+
\left\{
\gam^{(-)}_{\dwa}(\epl_+-\epl_0+\hfOmg)\cos^2 ({\textstyle \frac{\Theta}{2}})
+\gam^{(-)}_{\upa}(\epl_+-\epl_0-\hfOmg)\sin^2 ({\textstyle \frac{\Theta}{2}})\right\}
|u_+(0)\ra\la u_+(0)|\tilde{\rho}(\bmchi,t)
\nonumber \\
&+
\left\{\gam^{(-)}_{\dwa}(\epl_--\epl_0+\hfOmg) \sin^2 ({\textstyle \frac{\Theta}{2}})
+\gam^{(-)}_{\upa}(\epl_--\epl_0-\hfOmg) \cos^2 ({\textstyle \frac{\Theta}{2}})\right\}
|u_-(0)\ra\la u_-(0)|\tilde{\rho}(\bmchi,t)
\nonumber \\
&+
\frac{1}{2}\sin\Theta\; e^{+\rmi(\epl_+-\epl_-)t} \left\{\gam_\dwa^{(-)}(\epl_--\epl_0+\hfOmg)
-\gam_\upa^{(-)}(\epl_--\epl_0-\hfOmg)\right\}|u_+(0)\ra\la u_-(0)|\tilde{\rho}(\bmchi,t)
\nonumber \\
&+
\frac{1}{2}\sin\Theta\; e^{-\rmi(\epl_+-\epl_-)t}\left\{\gam_\dwa^{(-)}(\epl_+-\epl_0+\hfOmg)
-\gam_\upa^{(-)}(\epl_+-\epl_0-\hfOmg)\right\}|u_-(0)\ra\la u_+(0)|\tilde{\rho}(\bmchi,t).
\end{align}
\end{widetext}
The term [IV] in \Eq{B6} can be analyzed in a similar way.

We have introduced
\begin{align}\label{A7}
\gamma_\sgm^{(\pm)}(\omg)
=\sum_k \int_{0}^\infty \rmd \tau e^{-\rmi \omg \tau} C_\sgm^{(\pm)}(\tau),
\end{align}
where $C_\sgm^{(\pm)}(\tau)$ are the reservoir correlation functions
defined as
\bsube\label{A8}
\begin{align}
C_\sgm^{(+)}(\tau)&=\la \tilde{f}_\sgm^\dag(\tau) \tilde{f}_\sgm(0) \ra_\rmB,
\\
C_\sgm^{(-)}(\tau)&=\la \tilde{f}_\sgm(\tau) \tilde{f}_\sgm^\dag(0) \ra_\rmB,
\end{align}
\esube
with $\tilde{f}_\sgm(\tau)\equiv\tilde{f}_\sgm(\bmchi={\bm 0},\tau)$ and
$\la(\cdots)\ra_\rmB\equiv\tr_\rmB[(\cdots)\rho_\rmB]$ the usual thermal
average.
By substituting \Eq{AW} into \Eq{A8}, the reservoir correlation functions
simplifies to
\be
C_\sgm^{(\pm)}(\tau)=\sum_k  |t_{k\sgm}|^2 f^{(\pm)}(\vpl_{k\sgm})
e^{\pm\rmi\vpl_{k\sgm}\tau}.
\ee

Actually, \Eq{A7} is a causality transformation, which can be decomposed
into spectral functions and dispersion functions as\cite{Yan05187}
\begin{align}\label{ZZ2}
\gamma_\sgm^{(\pm)}(\omg)=\Gam_\sgm^{(\pm)}(\omg)+\rmi D_\sgm^{(\pm)}(\omg).
\end{align}
The involved spectral functions are simply the Fourier transforms
of the corresponding reservoir correlation functions
\be\label{Gamsgm}
\Gam_\sgm^{(\pm)}(\omg)
=\int_{-\infty}^\infty \rmd \tau e^{-\rmi\omg \tau}
C_\sgm^{(\pm)}(\tau).
\ee
In the usual wide-band limit, it reduces to
\be
\Gam_\sgm^{(\pm)}(\omg)=\Gam_\sgm f^{(\pm)}(\omg),
\ee
where $\Gam_\sgm $ is the tunneling width, $f^{(+)}(\omg)$ is
the usual Fermi function, and $f^{(-)}(\omg)\equiv 1-f^{(+)}(\omg)$.
With the knowledge of the spectral functions, the dispersion
functions in \Eq{ZZ2} can
be obtained via the Kramers-Kronig Relation \cite{Yan05187,Xu029196}
\be\label{Dsgm}
D_\sgm^{(\pm)}(\omg)=-\frac{1}{\pi}{\cal P}\int_{-\infty}^{\infty} \rmd \omg' \frac{C_\sgm^{(\pm)}(\omg')}{\omg-\omg'},
\ee
where ${\cal P}$ denotes the principle value.
%%%
By introducing a Lorentzian cutoff
$J(\omg)=\frac{{\rm w}^2}{(\omg-\mu_0)^2+{\rm w}^2}$ centered
at $\omg=\mu_0$ and with bandwidth ${\rm w}$, the dispersion functions
can be evaluated
\be
D_\sgm^{(\pm)}(\omg)=\pm \frac{\Gam_\sgm}{\pi}
\left\{\ln\left(\frac{\beta {\rm w}}{2\pi}\right)-{\rm Re}\Psi\left[\frac{1}{2}+\frac{\rmi\beta}{2\pi}(\omg-\mu_0)\right]\right\},
\ee
where $\Psi(x)$  is the digamma function.
The dispersion functions normally account for the system-bath coupling-induced
energy renormalization, similar to the so--called Lamb shift.
It has been revealed that the energy renormalization have strong influence
on electron transport through QD
systems \cite{Wun05205319,Luo11145301,Luo1159}, Aharonov-Bohm
interferometer \cite{Kon02045316,Mar03195305}, quantum measurement
of solid-state qubit \cite{Luo09385801,Luo104904}.
Later, we will show that the dispersion functions in ESR pumping have
important contribution to an effective magnetic field.

It is noted that in \Eq{ZZ1A} the coefficients are independent of the
counting fields.
Mathematically, this is due to fact that for nonzero reservoir correlation
functions, one should only account for the thermal averages of
$\tilde{f}_\sgm(\bmchi)$ and its Hermitian conjugate with the same momentum and
spin, cf. \Eq{A8}.
Physically, this implies that the term [I] in \Eq{ZZ1A} is not directly
responsible for particle and energy transport. However, it does not necessarily
mean that they donot have any contribution. Actually, they may have
important roles to play via influencing the spin dynamics.
Simpliar analysis applies to the last term [IV] in \Eq{B6},
which is also $\bmchi$-independent.
This is no long the case for the second and third terms, i.e.
[II] and [III] in \Eq{B6}, which depend explicitly on the counting
fields.
For instance, the term [II] is given by
\begin{widetext}
\begin{align}
{\rm [II]}=&\left\{
	\gamma_{\dwa}^{(+)}(\epl_+-\epl_0+{\textstyle \frac{\Omg}{2}})
	\cos^2(\textstyle{\frac{\Theta}{2}})e^{-\rmi (\epl_+-\epl_0+{\textstyle \frac{\Omg}{2}})\chi_{2\dwa}-\rmi\chi_{1\dwa}}
	+\gamma_{\upa}^{(+)}(\epl_+-\epl_0-{\textstyle \frac{\Omg}{2}})
	\sin^2(\textstyle{\frac{\Theta}{2}})e^{-\rmi (\epl_+-\epl_0-{\textstyle \frac{\Omg}{2}})\chi_{2\upa}-\rmi\chi_{1\upa}}\right\}
	\nonumber \\
	&\hspace{0.9cm}\times|u_+(0)\ra\la u_0(0)|
	\tilde{\rho}(\bmchi,t)|u_0(0)\ra\la u_+(0)|
	\nonumber \\
	&+\left\{
	\gamma_{\dwa}^{(+)}(\epl_--\epl_0+{\textstyle \frac{\Omg}{2}})
	\sin^2(\textstyle{\frac{\Theta}{2}})e^{-\rmi (\epl_--\epl_0+{\textstyle \frac{\Omg}{2}})\chi_{2\dwa}-\rmi\chi_{1\dwa}}
	+\gamma_{\upa}^{(+)}(\epl_--\epl_0-{\textstyle \frac{\Omg}{2}})
	\cos^2(\textstyle{\frac{\Theta}{2}})e^{-\rmi (\epl_--\epl_0-{\textstyle \frac{\Omg}{2}})\chi_{2\upa}-\rmi\chi_{1\upa}}\right\}
	\nonumber \\
	&\hspace{0.9cm}\times|u_-(0)\ra\la u_0(0)|
\tilde{\rho}(\bmchi,t)|u_0(0)\ra\la u_-(0)|
\nonumber \\
	&+\left\{\gamma_{\dwa}^{(-)}(\epl_+-\epl_0+{\textstyle \frac{\Omg}{2}})
	\cos^2(\textstyle{\frac{\Theta}{2}})e^{+\rmi (\epl_+-\epl_0+{\textstyle \frac{\Omg}{2}})\chi_{2\dwa}+\rmi\chi_{1\dwa}}+\gamma_{\upa}^{(-)}(\epl_+-\epl_0-{\textstyle \frac{\Omg}{2}})
	\sin^2(\textstyle{\frac{\Theta}{2}})e^{+\rmi (\epl_+-\epl_0-{\textstyle \frac{\Omg}{2}})\chi_{2\upa}+\rmi\chi_{1\upa}}\right\}
	\nonumber \\
	&\hspace{0.9cm}\times|u_0(0)\ra\la u_+(0)|
\tilde{\rho}(\bmchi,t)|u_+(0)\ra\la u_0(0)|
	\nonumber \\
	&+\left\{\gamma_{\dwa}^{(-)}(\epl_--\epl_0+{\textstyle \frac{\Omg}{2}})
	\sin^2(\textstyle{\frac{\Theta}{2}})e^{+\rmi (\epl_--\epl_0+{\textstyle \frac{\Omg}{2}})\chi_{2\dwa}+\rmi\chi_{1\dwa}}
	+\gamma_{\upa}^{(-)}(\epl_--\epl_0-{\textstyle \frac{\Omg}{2}})
	\cos^2(\textstyle{\frac{\Theta}{2}})e^{+\rmi (\epl_--\epl_0-{\textstyle \frac{\Omg}{2}})\chi_{2\upa}+\rmi\chi_{1\upa}}\right\}
	\nonumber \\
	&\hspace{0.9cm}\times|u_0(0)\ra\la u_-(0)|
\tilde{\rho}(\bmchi,t)|u_-(0)\ra\la u_0(0)|
\nonumber \\
&+\frac{1}{2}\sin\Theta \, e^{+\rmi (\epl_+-\epl_-)t}\left\{
\gamma_{\dwa}^{(+)}(\epl_+-\epl_0+{\textstyle \frac{\Omg}{2}})
e^{-\rmi (\epl_+-\epl_0+{\textstyle \frac{\Omg}{2}})\chi_{2\dwa}-\rmi\chi_{1\dwa}}
-\gamma_{\upa}^{(+)}(\epl_+-\epl_0-{\textstyle \frac{\Omg}{2}})
e^{-\rmi (\epl_+-\epl_0-{\textstyle \frac{\Omg}{2}})\chi_{2\upa}-\rmi\chi_{1\upa}}\right\}
\nonumber \\
&\hspace{0.9cm}\times|u_+(0)\ra\la u_0(0)|\tilde{\rho}(\bmchi,t)|u_0(0)\ra\la u_-(0)|
\nonumber \\
&+\frac{1}{2}\sin\Theta \, e^{-\rmi (\epl_+-\epl_-)t}\left\{
\gamma_{\dwa}^{(+)}(\epl_--\epl_0+{\textstyle \frac{\Omg}{2}})
e^{-\rmi (\epl_--\epl_0+{\textstyle \frac{\Omg}{2}})\chi_{2\dwa}-\rmi\chi_{1\dwa}}
-\gamma_{\upa}^{(+)}(\epl_--\epl_0-{\textstyle \frac{\Omg}{2}})
e^{-\rmi (\epl_--\epl_0-{\textstyle \frac{\Omg}{2}})\chi_{2\upa}-\rmi\chi_{1\upa}}\right\}
\nonumber \\
&\hspace{0.9cm}\times|u_-(0)\ra\la u_0(0)|\tilde{\rho}(\bmchi,t)|u_0(0)\ra\la u_+(0)|
\nonumber \\
&+\frac{1}{2}\sin\Theta \, e^{-\rmi (\epl_+-\epl_-)t}\left\{
\gamma_{\dwa}^{(-)}(\epl_+-\epl_0+{\textstyle \frac{\Omg}{2}})
e^{+\rmi (\epl_+-\epl_0+{\textstyle \frac{\Omg}{2}})\chi_{2\dwa}+\rmi\chi_{1\dwa}}
-\gamma_{\upa}^{(-)}(\epl_+-\epl_0-{\textstyle \frac{\Omg}{2}})
e^{+\rmi (\epl_+-\epl_0-{\textstyle \frac{\Omg}{2}})\chi_{2\upa}+\rmi\chi_{1\upa}}\right\}
\nonumber \\
&\hspace{0.9cm}\times|u_0(0)\ra\la u_+(0)|\tilde{\rho}(\bmchi,t)|u_-(0)\ra\la u_0(0)|
\nonumber \\
&+\frac{1}{2}\sin\Theta \, e^{+\rmi (\epl_+-\epl_-)t}\left\{
\gamma_{\dwa}^{(-)}(\epl_--\epl_0+{\textstyle \frac{\Omg}{2}})
e^{+\rmi (\epl_--\epl_0+{\textstyle \frac{\Omg}{2}})\chi_{2\dwa}+\rmi\chi_{1\dwa}}
-\gamma_{\upa}^{(-)}(\epl_--\epl_0-{\textstyle \frac{\Omg}{2}})
e^{+\rmi (\epl_--\epl_0-{\textstyle \frac{\Omg}{2}})\chi_{2\upa}+\rmi\chi_{1\upa}}\right\}
\nonumber \\
&\hspace{0.9cm}\times|u_0(0)\ra\la u_-(0)|\tilde{\rho}(\bmchi,t)|u_+(0)\ra\la u_0(0)|.
\end{align}
Accordingly, the term [III] can be obtained following the similar procedure.
By putting all the four terms in \Eq{B6} together, one finally arrive at the GQME in the
interaction picture as
\begin{align}\label{ZZCA}
\frac{\rmd}{\rmd t}\tilde{\rho}(\bmchi,t)
	=&\bigg\{\Gamma_{0,+}(\bmchi){\cal J}[|u_0(0)\ra\la u_+(0)|]
	-\frac{1}{2}\Gamma_{0,+}{\cal A}[|u_0(0)\ra\la u_+(0)|]
	-\frac{\rmi}{2}\kappa_{0,+}{\cal C}[|u_0(0)\ra\la u_+(0)|]	
	\bigg\}\tilde{\rho}(\bmchi,t)
	\nonumber \\
%%%
	&+\bigg\{\Gamma_{0,-}(\bmchi){\cal J}[|u_0(0)\ra\la u_-(0)|]
	-\frac{1}{2}\Gamma_{0,-}{\cal A}[|u_0(0)\ra\la u_-(0)|]
	-\frac{\rmi}{2}\kappa_{0,-}{\cal C}[|u_0(0)\ra\la u_-(0)|]	
	\bigg\}\tilde{\rho}(\bmchi,t)
	\nonumber \\
%%%
	&+\bigg\{\Gamma_{+,0}(\bmchi){\cal J}[|u_+(0)\ra\la u_0(0)|]
	-\frac{1}{2}\Gamma_{+,0}{\cal A}[|u_+(0)\ra\la u_0(0)|]
	-\frac{\rmi}{2}\kappa_{+,0}{\cal C}[|u_+(0)\ra\la u_0(0)|]		
	\bigg\}\tilde{\rho}(\bmchi,t)
	\nonumber \\
%%%
	&+\bigg\{\Gamma_{-,0}(\bmchi){\cal J}[|u_-(0)\ra\la u_0(0)|]
	-\frac{1}{2}\Gamma_{-,0}{\cal A}[|u_-(0)\ra\la u_0(0)|]
	-\frac{\rmi}{2}\kappa_{-,0}{\cal C}[|u_-(0)\ra\la u_0(0)|]
	\bigg\}\tilde{\rho}(\bmchi,t)
	\nonumber \\
%%%
&-e^{+\rmi(\epl_+-\epl_-)t}\{[\Upsilon_{0,-}+\rmi \xi_{0,-}]
|u_+(0)\ra\la u_-(0)| \tilde{\rho}(\bmchi,t)
+[\Upsilon_{0,+}-\rmi \xi_{0,+}] \tilde{\rho}(\bmchi,t)
|u_+(0)\ra\la u_-(0)|\}
\nonumber \\
&-e^{-\rmi(\epl_+-\epl_-)t}\{[\Upsilon_{0,+}+\rmi \xi_{0,+}]
|u_-(0)\ra\la u_+(0)|\tilde{\rho}(\bmchi,t)
+[\Upsilon_{0,-}-\rmi \xi_{0,-}]
\tilde{\rho}(\bmchi,t) |u_-(0)\ra\la u_+(0)|\}
\nonumber \\
&+e^{+\rmi(\epl_+-\epl_-)t}
[\Upsilon_{+,0}(\bmchi)+\rmi \xi_{+,0}(\bmchi)
+\Upsilon_{-,0}(\bmchi)-\rmi \xi_{-,0}(\bmchi)]
|u_+(0)\ra\la u_0(0)|\tilde{\rho}(\bmchi,t)	|u_0(0)\ra\la u_-(0)|
\nonumber \\
&+e^{+\rmi(\epl_+-\epl_-)t}
[\Upsilon_{0,+}(\bmchi)-\rmi \xi_{0,+}(\bmchi)
+\Upsilon_{0,-}(\bmchi)+\rmi \xi_{0,-}(\bmchi)] |u_0(0)\ra\la u_-(0)|\tilde{\rho}(\bmchi,t)	|u_+(0)\ra\la u_0(0)|
	\nonumber \\
&+e^{-\rmi(\epl_+-\epl_-)t}
[\Upsilon_{+,0}(\bmchi)-\rmi \xi_{+,0}(\bmchi)
+\Upsilon_{-,0}(\bmchi)+\rmi \xi_{-,0}(\bmchi)]
	|u_-(0)\ra\la u_0(0)|\tilde{\rho}(\bmchi,t)	|u_0(0)\ra\la u_+(0)|
	\nonumber \\	
&+e^{-\rmi(\epl_+-\epl_-)t}
[\Upsilon_{0,+}(\bmchi)+\rmi \xi_{0,+}(\bmchi)
+\Upsilon_{0,-}(\bmchi)-\rmi \xi_{0,-}(\bmchi)]
|u_0(0)\ra\la u_+(0)|\tilde{\rho}(\bmchi,t)|u_-(0)\ra\la u_0(0)|,	
\end{align}
\end{widetext}
where we have introduced the superoperators ${\cal J}[r]\rho=r\rho r^\dag$,
${\cal A}[r]\rho=r^\dag r \rho+\rho r^\dag r$,
and ${\cal C}[r]\rho=[r^\dag r,\rho]$.

The first four lines in \Eq{ZZCA} depict the tunneling between QD and side reservoir in the
Lindblad-like form, where the $\bmchi$-dependent terms are directly responsible
for particle and energy transfer. The involved $\bmchi$-dependent
rates are given by
\bsube\label{A23}
\begin{align}
	\Gamma_{0,\pm}(\bmchi)
	=&\Gamma^{(\dwa)}_{0,\pm}e^{+\rmi (\epl_\pm-\epl_0+{\textstyle \frac{\Omg}{2}})\chi_{2\dwa}+\rmi\chi_{1\dwa}}
	\nonumber \\
	&+\Gamma^{(\upa)}_{0,\pm}e^{+\rmi (\epl_\pm-\epl_0-{\textstyle \frac{\Omg}{2}})\chi_{2\upa}+\rmi\chi_{1\upa}},
	\\
	\Gamma_{\pm,0}(\bmchi)
	=&\Gamma^{(\dwa)}_{\pm,0}e^{-\rmi (\epl_\pm-\epl_0+{\textstyle \frac{\Omg}{2}})\chi_{2\dwa}-\rmi\chi_{1\dwa}}
	\nonumber \\
	&+\Gamma^{(\upa)}_{\pm,0}e^{-\rmi (\epl_\pm-\epl_0-{\textstyle \frac{\Omg}{2}})\chi_{2\upa}-\rmi\chi_{1\upa}},
\end{align}
\esube
with
\bsube\label{A24}
\begin{align}
	\Gamma^{(\dwa)}_{0,\pm}
	&=\frac{1\pm\cos\Theta}{2}
	\Gam_{\dwa}^{(-)}(\epl_{\pm}-\epl_0+{\textstyle \frac{\Omg}{2}}),
	\\
	\Gamma^{(\upa)}_{0,\pm}
	&=\frac{1\mp\cos\Theta}{2}
	\Gam_{\upa}^{(-)}(\epl_{\pm}-\epl_0-{\textstyle \frac{\Omg}{2}}),
	\\
	\Gamma^{(\dwa)}_{\pm,0}
	&=\frac{1\pm\cos\Theta}{2}
	\Gam_{\dwa}^{(+)}(\epl_{\pm}-\epl_0+{\textstyle \frac{\Omg}{2}}),
	\\
	\Gamma^{(\upa)}_{\pm,0}
	&=\frac{1\mp\cos\Theta}{2}
	\Gam_{\upa}^{(+)}(\epl_{\pm}-\epl_0-{\textstyle \frac{\Omg}{2}}).
\end{align}
\esube
The spectral functions $\Gam_{\sgm}^{(\pm)}(\omg)$
are given in \Eq{Gamsgm}.
The corresponding $\bmchi$-independent rates are simply obtained by setting $\bmchi={\bm 0}$, i.e., $\Gamma_{0,\pm}=\Gamma_{0,\pm}(\bmchi={\bm 0})$,
and likewise for $\Gamma_{\pm,0}$.
The terms $\kappa_{0,\pm}$ and $\kappa_{\pm,0}$ are only
related to energy renormalization and thus not involved in particle
and energy transport. They do not depend on the counting fields
\bsube\label{A25}
\begin{align}
	\kappa_{0,\pm}
	=&\frac{1\pm\cos\Theta}{2}
	D_{\dwa}^{(-)}(\epl_{\pm}-\epl_0+{\textstyle \frac{\Omg}{2}})
	\nonumber \\
	&+\frac{1\mp\cos\Theta}{2}
	D_{\upa}^{(-)}(\epl_{\pm}-\epl_0-{\textstyle \frac{\Omg}{2}}),
	\\
	\kappa_{\pm,0}
	=&\frac{1\pm\cos\Theta}{2}
	D_{\dwa}^{(+)}(\epl_{\pm}-\epl_0+{\textstyle \frac{\Omg}{2}})
	\nonumber \\
	&+\frac{1\mp\cos\Theta}{2}
	D_{\upa}^{(+)}(\epl_{\pm}-\epl_0-{\textstyle \frac{\Omg}{2}}).
\end{align}
\esube
The corresponding dispersion functions $D_{\sgm}^{\pm}(\omg)$
can be found in \Eq{Dsgm}.

All the terms in the last six lines of \Eq{ZZCA} are oscillating in time.
They  originate purely from the non-secular
treatment, where we also find some $\bmchi$-dependent terms. These terms
also have important roles to play in energy and particle exchange between the
QD and side reservoir. The $\bmchi$-dependent coefficients read
\bsube\label{A27}
\begin{align}
\Upsilon_{0,\pm}(\bmchi)
=&\Upsilon^{(\dwa)}_{0,\pm} e^{+\rmi (\epl_\pm-\epl_0+{\textstyle \frac{\Omg}{2}})\chi_{2\dwa}+\rmi\chi_{1\dwa}}
\nonumber \\
&-\Upsilon^{(\upa)}_{0,\pm}e^{+\rmi (\epl_\pm-\epl_0-{\textstyle \frac{\Omg}{2}})\chi_{2\upa}+\rmi\chi_{1\upa}},
\\
\Upsilon_{\pm,0}(\bmchi)
=&\Upsilon^{(\dwa)}_{\pm,0}e^{-\rmi (\epl_\pm-\epl_0+{\textstyle \frac{\Omg}{2}})\chi_{2\dwa}-\rmi\chi_{1\dwa}}
\nonumber \\
&-\Upsilon^{(\upa)}_{\pm,0}e^{-\rmi (\epl_\pm-\epl_0-{\textstyle \frac{\Omg}{2}})\chi_{2\upa}-\rmi\chi_{1\upa}},
\\
\xi_{0,\pm}(\bmchi)
=&\xi^{(\dwa)}_{0,\pm} e^{+\rmi (\epl_\pm-\epl_0+{\textstyle \frac{\Omg}{2}})\chi_{2\dwa}+\rmi\chi_{1\dwa}}
\nonumber \\
&-\xi^{(\upa)}_{0,\pm}e^{+\rmi (\epl_\pm-\epl_0-{\textstyle \frac{\Omg}{2}})\chi_{2\upa}+\rmi\chi_{1\upa}},
\\
\xi_{\pm,0}(\bmchi)
=&\xi^{(\dwa)}_{\pm,0}e^{-\rmi (\epl_\pm-\epl_0+{\textstyle \frac{\Omg}{2}})\chi_{2\dwa}-\rmi\chi_{1\dwa}}
\nonumber \\
&-\xi^{(\upa)}_{\pm,0}e^{-\rmi (\epl_\pm-\epl_0-{\textstyle \frac{\Omg}{2}})\chi_{2\upa}-\rmi\chi_{1\upa}},
\end{align}
\esube
where the individual spin-dependent components are given by
\bsube\label{A28}
\begin{align}
\Upsilon^{(\dwa)}_{0,\pm}
&=\frac{\sin\Theta}{4}
\Gam_{\dwa}^{(-)}(\epl_{\pm}-\epl_0+{\textstyle \frac{\Omg}{2}}),
\\
\Upsilon^{(\upa)}_{0,\pm}
&=\frac{\sin\Theta}{4}
\Gam_{\upa}^{(-)}(\epl_{\pm}-\epl_0-{\textstyle \frac{\Omg}{2}}),
\\
\Upsilon^{(\dwa)}_{\pm,0}
&=\frac{\sin\Theta}{4}
\Gam_{\dwa}^{(+)}(\epl_{\pm}-\epl_0+{\textstyle \frac{\Omg}{2}}),
\\
\Upsilon^{(\upa)}_{\pm,0}
&=\frac{\sin\Theta}{4}
\Gam_{\upa}^{(+)}(\epl_{\pm}-\epl_0-{\textstyle \frac{\Omg}{2}}),
\\
\xi^{(\dwa)}_{0,\pm}
&=\frac{\sin\Theta}{4}
D_{\dwa}^{(-)}(\epl_{\pm}-\epl_0+{\textstyle \frac{\Omg}{2}}),
\\
\xi^{(\upa)}_{0,\pm}
&=\frac{\sin\Theta}{4}
D_{\upa}^{(-)}(\epl_{\pm}-\epl_0-{\textstyle \frac{\Omg}{2}}),
\\
\xi^{(\dwa)}_{\pm,0}
&=\frac{\sin\Theta}{4}
D_{\dwa}^{(+)}(\epl_{\pm}-\epl_0+{\textstyle \frac{\Omg}{2}}),
\\
\xi^{(\upa)}_{\pm,0}
&=\frac{\sin\Theta}{4}
D_{\upa}^{(+)}(\epl_{\pm}-\epl_0-{\textstyle \frac{\Omg}{2}}).
\end{align}
\esube

In the case when all the terms in the last six lines of \Eq{ZZCA} are
oscillating fast, the effects of these terms will very rapidly average to zero.
It is then justified to apply the secular
approximation to drop these fast oscillating terms.
By further setting $\bmchi={\bm 0}$, one will arrive at a Lindblad quantum
master equation such that the populations and coherences are dynamically
decoupled \cite{Sil14022103,Cue15055002}.
All the thermodynamics can be analyzed in analogy to that for time-independent
situations.
By comparing the quasienergies $\epl_+$ and $\epl_-$ in \Eq{AV}, one
readily finds that fast oscillations only take place in the limit where the Rabi
frequency is much larger than the dissipation strength.

In this work, our investigation is based on the GQME beyond the secular
approximation, such that our thermodynamic analysis is valid for a wide
range of Rabi frequencies.
In particular, we will reveal the essential roles that the non--secular
treatment will play in the thermodynamics of the ESR pumped QD device.
Finally, by converting from the interaction picture back into
Schr\"odinger's picture, we arrives at the GQME in the Floquet
basis in \Eq{ZZC}.

%\bibliography{D:/bibliography/bibrefs}

\end{document}